\documentclass[letterpaper,12pt]{extarticle}  

\usepackage{import}
\usepackage{local}
\usepackage[left=.75in,top=.75in,bottom=.75in,right=.75in]{geometry}

\usepackage[utf8]{inputenc}
\usepackage[T1]{fontenc}

\usepackage{float}
\usepackage{physics}
\usepackage{graphicx}
\usepackage{dcolumn}
\usepackage{bm}
\usepackage{mathrsfs}
\usepackage{amssymb}
\usepackage{amsmath}
\usepackage[normalem]{ulem}
\usepackage{wrapfig}
\usepackage[compress,square,sort,comma,numbers]{natbib}

\usepackage{hyperref}
\usepackage{bookmark}

\hypersetup{%
  breaklinks=true,
  colorlinks=true,
  linkcolor=black,
  filecolor=black,
  urlcolor=black,
  citecolor=black,
  pdfborder=0 0 0,
}

\newcommand{\tnn}{\text{NN}}
\newcommand{\texp}{\text{exp}}

\begin{document}

\title{\LARGE Zyxin is all you need: machine learning adherent cell mechanics}

\author{%
\textbf{
Matthew S. Schmitt\textcolor{Accent}{\textsuperscript{1,2,3,*}}, %
Jonathan Colen\textcolor{Accent}{\textsuperscript{1,2,3,*}}, %
Stefano Sala\textcolor{Accent}{\textsuperscript{4}}, %
John Devany\textcolor{Accent}{\textsuperscript{1,2}}, %
Shailaja Seetharaman\textcolor{Accent}{\textsuperscript{1,2}}, %
 Margaret L. Gardel\textcolor{Accent}{\textsuperscript{1,2**}}, %
 Patrick W. Oakes\textcolor{Accent}{\textsuperscript{4**}}, %
 Vincenzo Vitelli\textcolor{Accent}{\textsuperscript{1,2,3**}} }\\
\begin{small}
\textcolor{Accent}{\textsuperscript{1}}James Franck Institute, University of Chicago, Chicago, Illinois 60637, U.S.A. \\ 
\textcolor{Accent}{\textsuperscript{2}}Department of Physics, University of Chicago, Chicago, IL 60637, U.S.A. \\
\textcolor{Accent}{\textsuperscript{3}}Kadanoff Center for Theoretical Physics, University of Chicago, Chicago, IL 60637, U.S.A. \\
\textcolor{Accent}{\textsuperscript{4}}Department of Cell \& Molecular Physiology, Stritch School of Medicine, Loyola University Chicago, Maywood, IL 60153, U.S.A. \\
\textcolor{Accent}{\textsuperscript{*}} These authors contributed equally \\ 
\textcolor{Accent}{\textsuperscript{**}}Correspondence: \textcolor{Accent}{gardel@uchicago.edu, poakes@luc.edu, vitelli@uchicago.edu} \\ \end{small}
}

\maketitle

\begin{onehalfspacing}

\section{abstract}
Cellular form and function emerge from complex mechanochemical systems within the cytoplasm.
No systematic strategy currently exists to infer large-scale physical properties of a cell from its many molecular components. This is a significant obstacle to understanding biophysical processes such as cell adhesion and migration.
Here, we develop a data-driven biophysical modeling approach to learn the mechanical behavior of adherent cells. 
We first train neural networks to predict forces generated by adherent cells from images of cytoskeletal proteins.
Strikingly, experimental images of a single focal adhesion protein, such as zyxin, are sufficient to predict forces and generalize to unseen biological regimes. 
This protein field alone contains enough information to yield accurate predictions even if forces themselves are generated by many interacting proteins.
We next develop two approaches -- one explicitly constrained by physics, the other more agnostic -- that help construct data-driven continuum models of cellular forces using this single focal adhesion field.
Both strategies consistently reveal that cellular forces are encoded by two different length scales in adhesion protein distributions. 
Beyond adherent cell mechanics, our work serves as a case study for how to integrate neural networks in the construction of predictive phenomenological models in cell biology, even when little knowledge of the underlying microscopic mechanisms exist.

\newpage
\section{Introduction}

The structure and dynamics of living cells are controlled by the physical properties of the cytoskeleton~\cite{Pegoraro2017,Blanchoin2014}. 
The cytoskeleton itself, however, is the product of complex biochemical circuits which regulate its dynamics and spatial organization~\cite{Fletcher2010,Svitkina2018}. 
The central challenge faced when studying the physical biology of the cell is to untangle this interplay between mechanical and biochemical constraints. 
Current modeling approaches have leaned heavily on intuition built upon centuries of work on classical continuum mechanics, where symmetries and conservation laws dictate both the variables which arise in such models as well as the equations they obey~\cite{Phillips2009}. 
Cells, however, are decidedly non-classical, relying instead on distributed enzymatic activity, non-equilibrium dynamics, and hierarchical structures~\cite{MacKintosh2010, Battle2016}. 
All of these features complicate coarse-graining and system parameterization in terms of a few collective simply-understood variables~\cite{Prost2015,Romani2021}. 
Thus, despite significant advancements and efforts in the field of physical biology, we still cannot even accurately predict the mechanical response of cells to biochemical perturbations. 

Machine learning (ML) has the potential to bypass some of the challenges faced by classical modeling approaches by discovering models directly from the statistics of data~\cite{Carleo2019, Cichos2020, Zaritsky2021, Soelistyo2022}. 
This approach has proven very successful in structural biology where recent machine-learning advances have largely solved the long-standing problem of predicting protein structures from gene sequences~\cite{Jumper2021,Lin2022biorxiv}.
In this work, we explore the potential of ML approaches to generate models of adherent cell mechanics.
Contractile forces generated by adherent cells are critical regulators of cell shape, adhesion, motility, and mechanotransduction~\cite{Iskratsch2014,Murrell2015}. 
Forces generated in the actin cytoskeleton are transmitted via transmembrane focal adhesions (FAs) to the extracellular matrix \cite{Schwarz2012, Burridge2015} where they can be measured directly with techniques like Traction Force Microscopy (TFM) \cite{Sabass2008, Huang2019}. 
TFM measurements coupled with live cell imaging of fluorescently-tagged cytoskeletal proteins have helped develop a number of biophysical models of cellular force generation and mechanosensing~\cite{mertz2012, oakes2014, Soine2015, Cao2015, Notbohm2016, Oakes2017,  Hanke2018,Vignaud2021}. 
While providing insight into various local microscopic mechanisms, these models do not capture the broad heterogeneity of structures and behaviors in cells, and cannot fully account for how non-local and cell-scale properties such as cell morphology and FA location affect, and even dominate, local force measurements.

\begin{figure}
    \centering
    \includegraphics[width=.95\textwidth]{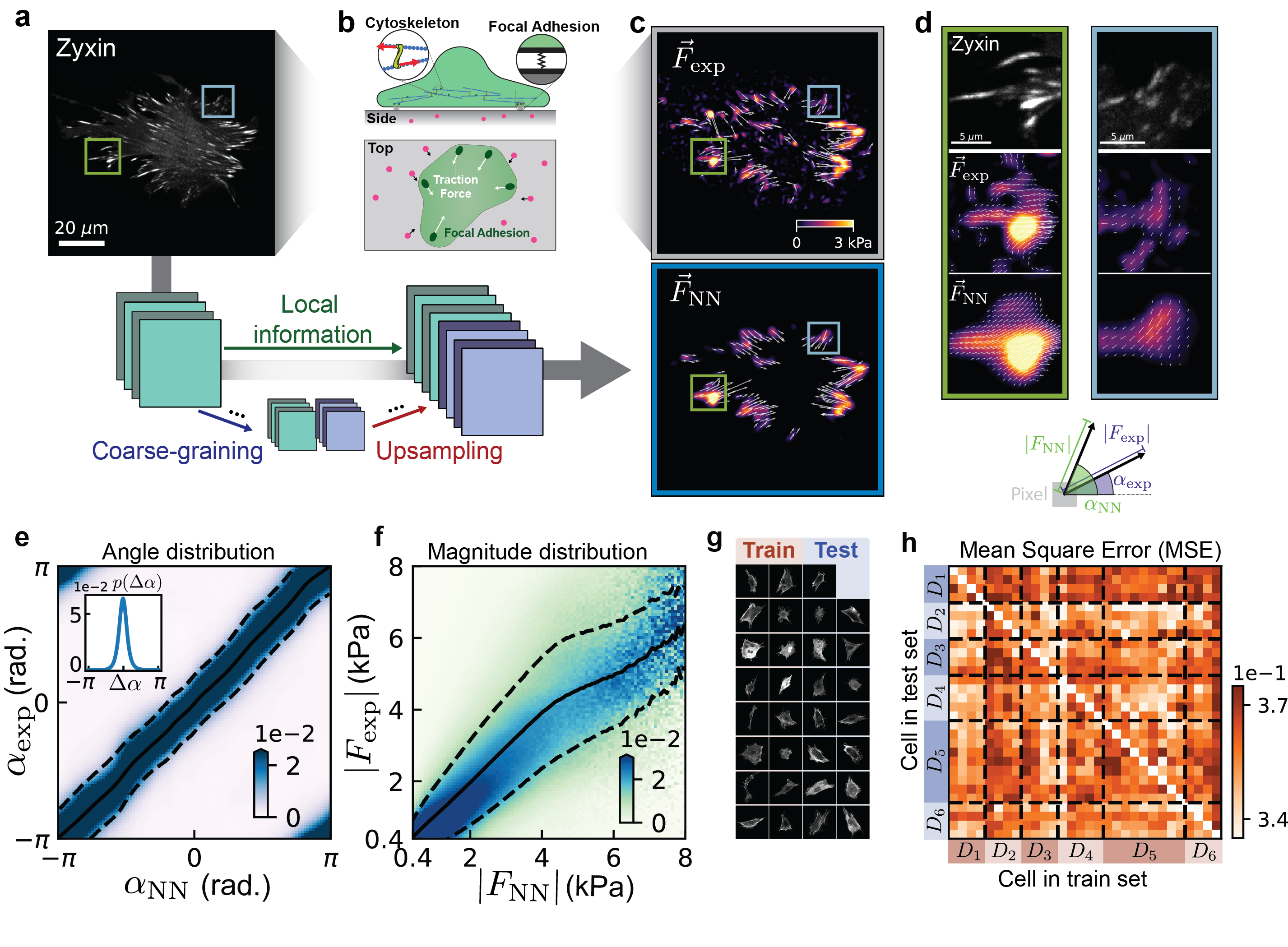}
    \caption{\textbf{Neural networks accurately predict traction forces.} \\
    (\textbf{a}) Fluorescent protein intensities (e.g. EGFP-zyxin) are measured in cells spread on 2D polyacrylamide gels coated with fibronectin. 
    (\textbf{b}) Adherent cells generate forces via the contractile activity of the cytoskeleton. These traction forces are transmitted to the substrate through focal adhesions (FAs). By measuring the displacement of fluorescent beads embedded in the substrate (red circles), the traction forces can be reconstructed using Traction Force Microscopy (TFM; see Methods). 
    (\textbf{c}) (Top) Forces ($\vec{F}_\text{exp}$) recovered from experimental measurements of substrate deformations via TFM. (Bottom) U-Nets predict traction forces ($\vec{F}_\text{NN}$) from images of protein intensity. In both plots the magnitude of the traction force is indicated by the color, and the direction by the overlaid arrow. 
    (\textbf{d}) Zoomed-in view of colored boxes in (\textbf{c}). 
    (\textbf{e}) At each pixel, we measure $(\alpha_{\text{exp}}, \alpha_{\text{NN}},|F_{\text{exp}}|,|F_{\text{NN}}|)$, which we bin to calculate the conditional angular distribution $p(\alpha_{\text{exp}}|\alpha_{\text{NN}})$ (\textbf{e}) and (\textbf{f}) the conditional magnitude distribution $p(|F_{\text{exp}}|\big||F_{\text{NN}}|)$. An optimal predictor lies exactly along the diagonal. Solid lines denote the average of the distribution while dashed lines mark one standard deviation. The angular distribution is strongly peaked along this diagonal (with additional peaks appearing due to periodicity), while the magnitude distribution remains on the diagonal up to $|F_{\text{exp}}|\approx 4$ kPa, which corresponds to 99.9\% of pixels.
    Inset of (\textbf{e}) shows the probability distribution of angular error $\Delta\alpha=\alpha_\text{NN} - \alpha_\text{exp}$.
    (\textbf{g})  Partition of 31-cell dataset into 16-cell training set and 15-cell test set. Every cell shown in this paper is in the test set and was not seen during training.
    (\textbf{h}) Model mean-square error for 22 random train/test partitions. Dashed lines denote days on which cells were imaged. Pixel color $p_{ij}$ is the average MSE of all models which use cell $i$ for training and cell $j$ for testing. }
    \label{fig:unet_predictions}
\end{figure}

In this work, we demonstrate how the flexibility of neural networks can be harnessed to both improve existing models as well as to discover entirely new ones.
We begin by training deep neural networks to predict forces directly from images of fluorescent cytoskeletal proteins, and in the process identify a minimal set of FA proteins which are sufficient to predict traction stresses.
We show that this network's predictions generalize to both previously unseen experimental and biological perturbations, and then probe the network to identify features which inform the network's predictions. 
Next, we use a physics-constrained ML approach to build on existing mechanical cellular models~\cite{edwards2011prl, mertz2012, oakes2014, solowiej-wedderburn2022} by learning a mapping between measured protein distributions and the physically-meaningful parameters in a linear-elastic model.
Finally, by casting away our mechanical hypotheses we demonstrate a purely data-driven pipeline which constructs relevant fields and distills a set of effective equations which predict cellular traction stresses.
Despite varying degrees of model complexity and prior knowledge, all of our approaches consistently reveal that models for force generation are characterized by the interaction of both local and non-local features.
These findings illustrate how FA proteins encode information of local forces at adhesion sites as well as whole-cell contractility through their distribution in the cell, and demonstrate a suite of complementary approaches to build novel models of living systems. 

\section{Results}
\subsection*{Neural networks predict traction forces from images of proteins}
To assess whether neural networks could make mechanical predictions from biochemical fields, we  created a library comprised of paired images of the FA protein zyxin in fibroblasts (\cite{Hoffman2006}; Fig.~\ref{fig:unet_predictions}a) with their corresponding traction forces as directly measured by TFM ($\vec{F}_{\text{exp}}$; Fig.~\ref{fig:unet_predictions}b-c).
As expected, traction forces primarily localized along the cell boundary at FAs as marked by zyxin accumulation, and pointed inwards towards the cell body (Fig.~\ref{fig:unet_predictions}c,d). 
For our neural network we chose a U-Net architecture which learns large-scale features via successive strided convolutions, while skip connections between layers propagate fine-grained information and preserve local structure that may be lost during coarse-graining (\cite{ronneberger2015unet}; Fig.~\ref{fig:unet_predictions}a).
Using the library of zyxin images as inputs, the U-Net was trained to directly predict traction forces ($\vec{F}_{\text{NN}}$; Fig.~\ref{fig:unet_predictions}c,d). 
In total our library contained 3,720 images, obtained from 31 separate time-lapses of cells expressing zyxin and their associated traction force fields. 
The library was split into training and test sets containing 16 and 15 cells, respectively, with the training data used to teach the U-Net and the test data used to evaluate the network predictions (Fig.~\ref{fig:unet_predictions}g). 
Following training, the U-Net was able to make accurate predictions of traction forces for cells from the test set which the network had never seen (Fig.~\ref{fig:unet_predictions}c,d).
The network predictions of traction forces agreed generally with experimental measurements in both location and magnitude (Fig.~\ref{fig:unet_predictions}c), with some smoothing occurring at the micron scale (Fig.~\ref{fig:unet_predictions}d).

\begin{figure*}
    \centering
    \includegraphics[width=.9\textwidth]{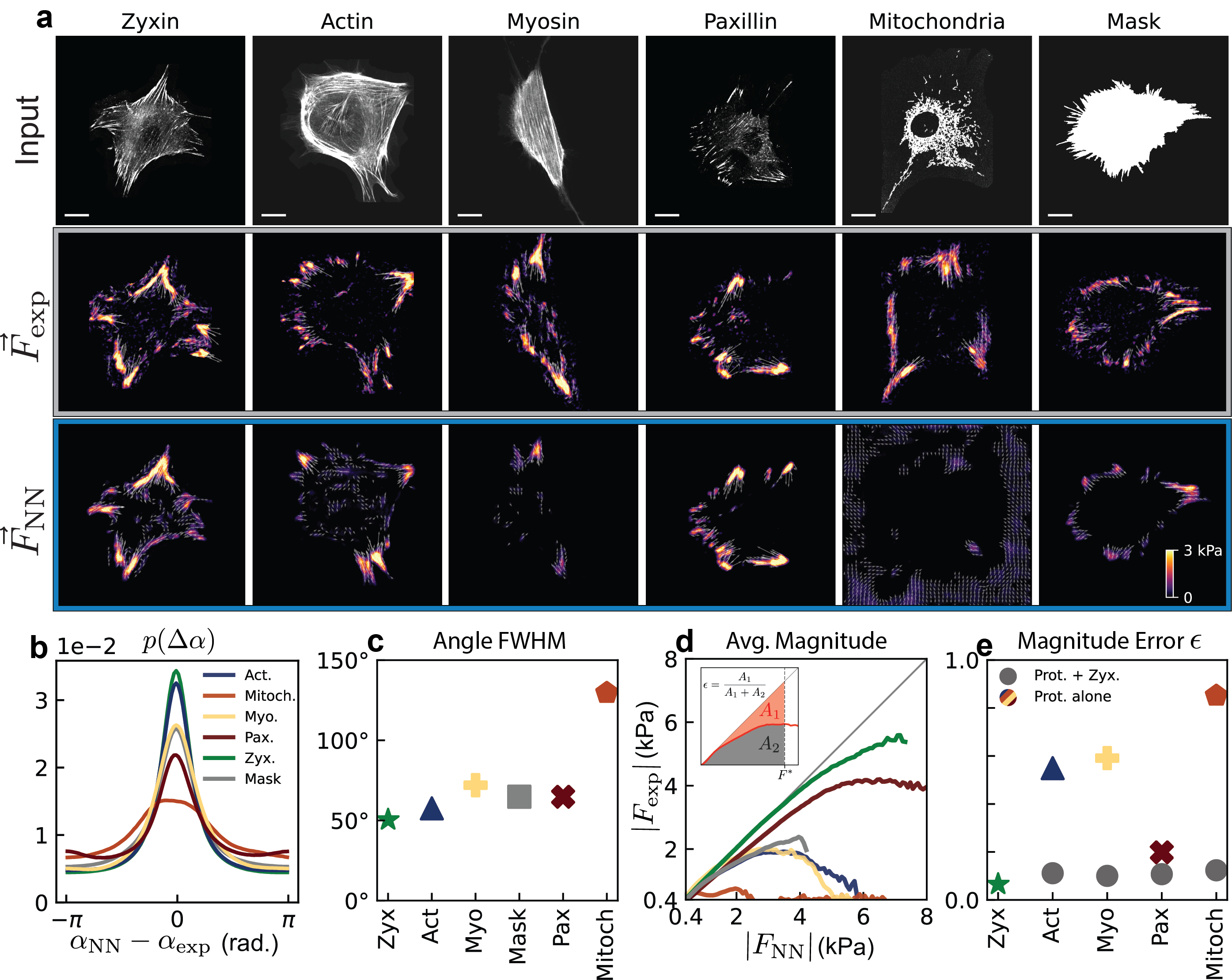}
    \caption{\textbf{Networks trained on zyxin outperform those trained on other proteins}
    \\
    (\textbf{a}) The predictive power of different cellular proteins inputs are compared by training neural networks on each protein individually. The comparison includes cytoskeletal proteins associated with force generation (actin, myosin), FA proteins (paxillin, zyxin), as well as a protein thought to be unrelated to force generation (mitochondria) and the binary cell mask.
    (\textbf{b}) While all networks trained in (\textbf{a}) predict accurate force directions on average, the distribution of errors varies depending on protein.
    (\textbf{c}) We quantify the angular error by the full width half maximum (FWHM) value of the distributions in (\textbf{b}). The networks performed similarly, with the exception of the mitochondria network which showed a much larger FWHM.
    (\textbf{d}) NNs trained on focal adhesion proteins, in particular zyxin, predict force magnitudes more accurately than those trained on other inputs.
    Inset shows calculation of magnitude error, which measures the cumulative distance from the diagonal up to $F^*=6$ (black dashed line). 
    (\textbf{e}) Zyxin outperforms all other proteins in predicting force magnitudes, and training on zyxin plus other proteins does not improve performance. 
    }
    \label{fig:proteins}
\end{figure*}

To evaluate the U-Net predictions, we compared the experimentally measured traction force directions ($\alpha_{\text{exp}}$) and magnitudes ($|F_{\text{exp}}|$) to those predicted by the U-Net for all the cells in the test set ($\alpha_{\text{NN}}$, $|F_{\text{NN}}|$).
Fig.~\ref{fig:unet_predictions}e,f shows the conditional distributions $p(\alpha_{\text{exp}}\big|\alpha_{\text{NN}})$ and $p(|F_{\text{exp}}|\big||F_{\text{NN}}|)$ (see Methods for additional details) along with the averages (solid line) and standard deviation (dotted line).
The neural network achieves near-optimal accuracy for force angles as well as magnitudes up to $\sim$4~kPa, which represents approximately 99.95\% of all the pixels in the test dataset (Fig.~\ref{fig:unet_predictions}e,f).
To evaluate the neural network's sensitivity to the test data used, we generated 22 random partitions of our 31-cell library into 16-cell training sets and 15-cell test sets.
We trained a separate U-Net on each partition, 
and evaluated the mean square error (MSE) of the force predictions. 
This revealed no systematic variations or dependencies on the choice of cells in the training set. 
Instead, the variations in accuracy caused by different training cells were no greater than the MSE fluctuations between experimental days, remaining within 10\% of the average MSE (Fig.~\ref{fig:unet_predictions}h).
These results demonstrate that from a readily achievable amount of experimental data, a U-Net can robustly learn to make accurate predictions of traction forces from fluorescent images of a focal adhesion protein. 

\subsection*{Zyxin is more predictive than other proteins}

In addition to identifying FAs, zyxin also reveals information about actin stress fiber organization and general cell geometry \cite{Yoshigi2005-ri}.
To determine which of these features was driving the U-Net performance, we tested the efficacy of other cytoskeletal proteins involved in force transmission, including: actin and myosin, the filaments and motors which make up the contractile network; paxillin, another focal adhesion protein; mitochondria, an organelle unconnected to the contractile machinery as a negative control; and binary masks of the cell morphology. 
For these experiments, we simultaneously expressed zyxin with the other proteins of interest. 
With the exception of mitochondria-trained networks, all networks learned to predict forces to some degree of accuracy, capturing the general localization and magnitude of traction stresses (Fig.~\ref{fig:proteins}a).
The probability distribution of angular error $\Delta\alpha=\alpha_{\text{NN}} - \alpha_{\text{exp}}$ peaked around zero for all proteins, differing only in the width of the distribution about the true value (Fig.~\ref{fig:proteins}b).
This distribution width was similar for networks trained on each protein, with the exception of mitochondria which showed a high angular variance.
When comparing force magnitude predictions, we observed larger differences amongst the proteins, with the FA proteins zyxin and paxillin outperforming all others
(Fig.~\ref{fig:proteins}d). 
Surprisingly, training networks on combined inputs of zyxin and these proteins did not improve performance, and performed as well as zyxin alone (Fig.~\ref{fig:proteins}e,~SI~Fig~5).
These results demonstrate that neural networks can be used to sort through potential proteins and identify a minimal subset which contains all the necessary information about the cell to predict forces.
We proceed using our highest-performing neural network, which was trained using zyxin alone.

\subsection*{Zyxin-trained networks generalize to new cell types and biological perturbations}
\begin{figure}
    \centering
    \includegraphics[width=0.9\textwidth]{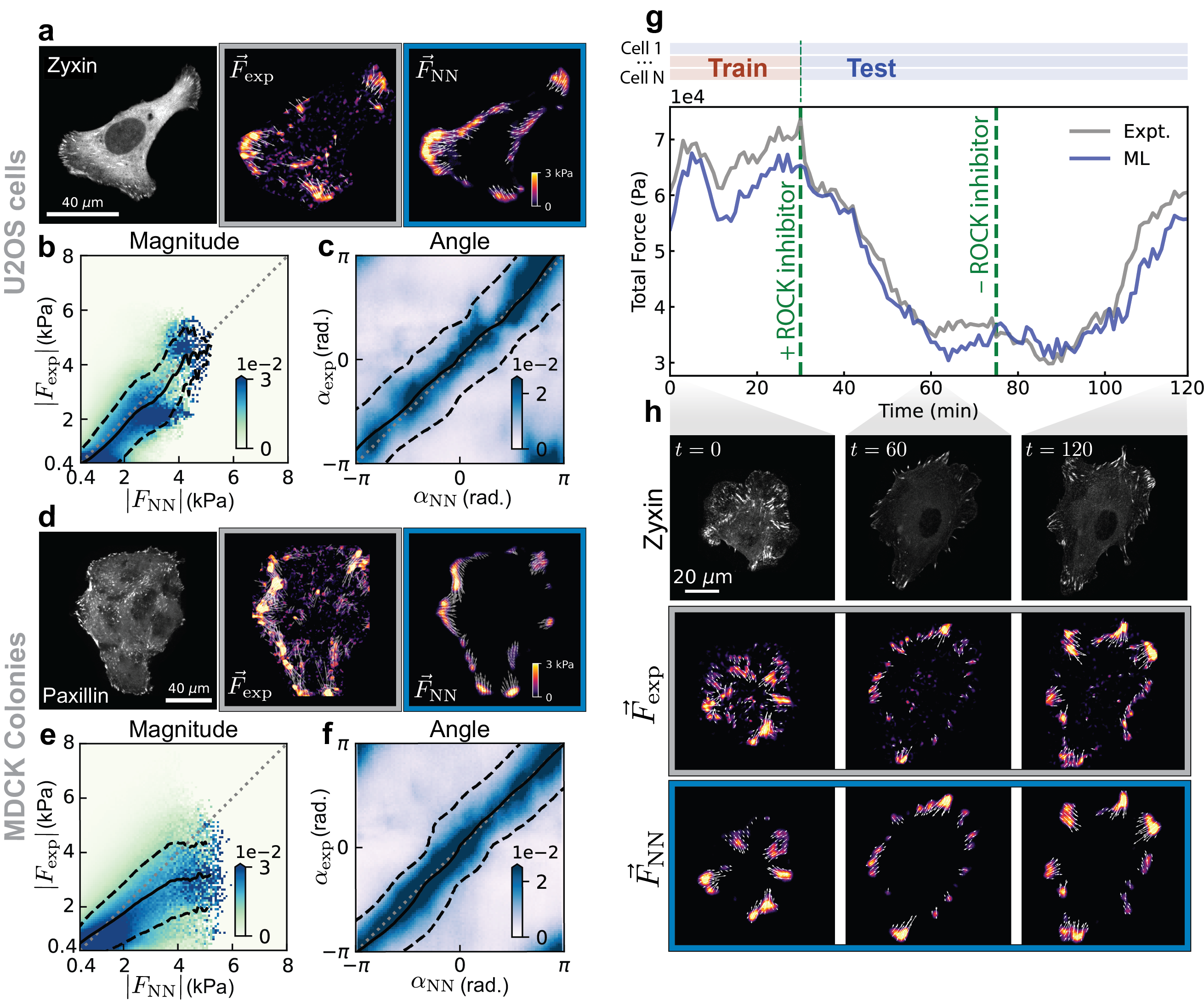}
    \caption{\textbf{Neural networks generalize across cell types and biomechanical regimes}
    \\
    (\textbf{a-c}) Networks trained on fibroblasts are evaluated on (\textbf{e}) individual U2OS cells expressing zyxin, and (\textbf{f}) colonies of MDCK cells expressing paxillin. 
    (\textbf{g-j}) Pixel-wise distributions of angle and magnitude predictions for U2OS cells (\textbf{g-h}) and MDCK cells (\textbf{i-j}), as in Fig.~\ref{fig:unet_predictions}. 
    (\textbf{k}) A neural network is trained on fibroblasts in their basal contractile state and evaluated on fibroblasts perturbed with the ROCK inhibitor Y-27632. The wash-in at $t=30$min impairs cytoskeletal contractility resulting in lower total force, which recovers after the drug is washed out at $t=75$min. 
    (\textbf{l}) Three snapshots from the time series in (\textbf{k}) demonstrate the NN's ability to capture redistributions of forces seen during the perturbation. 
    }
    \label{fig:generalization}
\end{figure}

While it is generally assumed that the underlying mechanics of contraction are universal~\cite{Murrell2015}, we sought to explicitly test this by evaluating our U-Net (which was trained on images of fibroblasts) on images of other adherent cell types. 
Specifically, we imaged zyxin in individual human osteosarcoma epithelial cells (U2OS; Fig.~\ref{fig:generalization}a) and paxillin in colonies of canine epithelial cells (MDCK; Fig.~\ref{fig:generalization}d). 
Without any retraining, the zyxin-trained U-Net generally predicted accurate traction force directions and magnitudes for both new cell types (Fig.~\ref{fig:generalization}b-c,e-f).
The ability of the network to generalize to different cell types, adhesion proteins, and cell clusters suggests that it has learned some underlying fundamental laws governing traction force generation.

To probe this idea further, we next challenged our U-Net model to make predictions in response to a biochemical perturbation. 
We imaged cells for 30 minutes at a basal contractile state before adding 5$~\mu$M of the Rho Kinase (ROCK) inhibitor Y-27632 for 45 minutes, and then washing out the drug and imaging for a final 45 minutes (Fig.~\ref{fig:generalization}g-h). 
Adding Y-27632 resulted in a drop in traction forces, an increase in overall cell area, and a reduction in the size of FAs, as expected~\cite{stricker2013,Oakes2018}, while the washout reversed each of these trends. 
To ensure that we tested the U-Net's ability to generalize and not memorize, we only trained the network using frames from the basal state (i.e. the first 30 minutes of imaging, prior to any drug treatment) of experiments. 
Despite having never seen these drug perturbations, the network still predicted the overall changes in global traction forces (Fig.~\ref{fig:generalization}g) and the local changes at FAs (Fig.~\ref{fig:generalization}h) during both the drug treatment and the subsequent recovery following wash-out. 
Together, these results indicate that the distribution of zyxin is a faithful proxy for the mechanical state of a cell and is sufficient to predict traction forces under a wide variety of conditions. 

\begin{figure*}
    \centering
    \includegraphics[width=1.\textwidth]{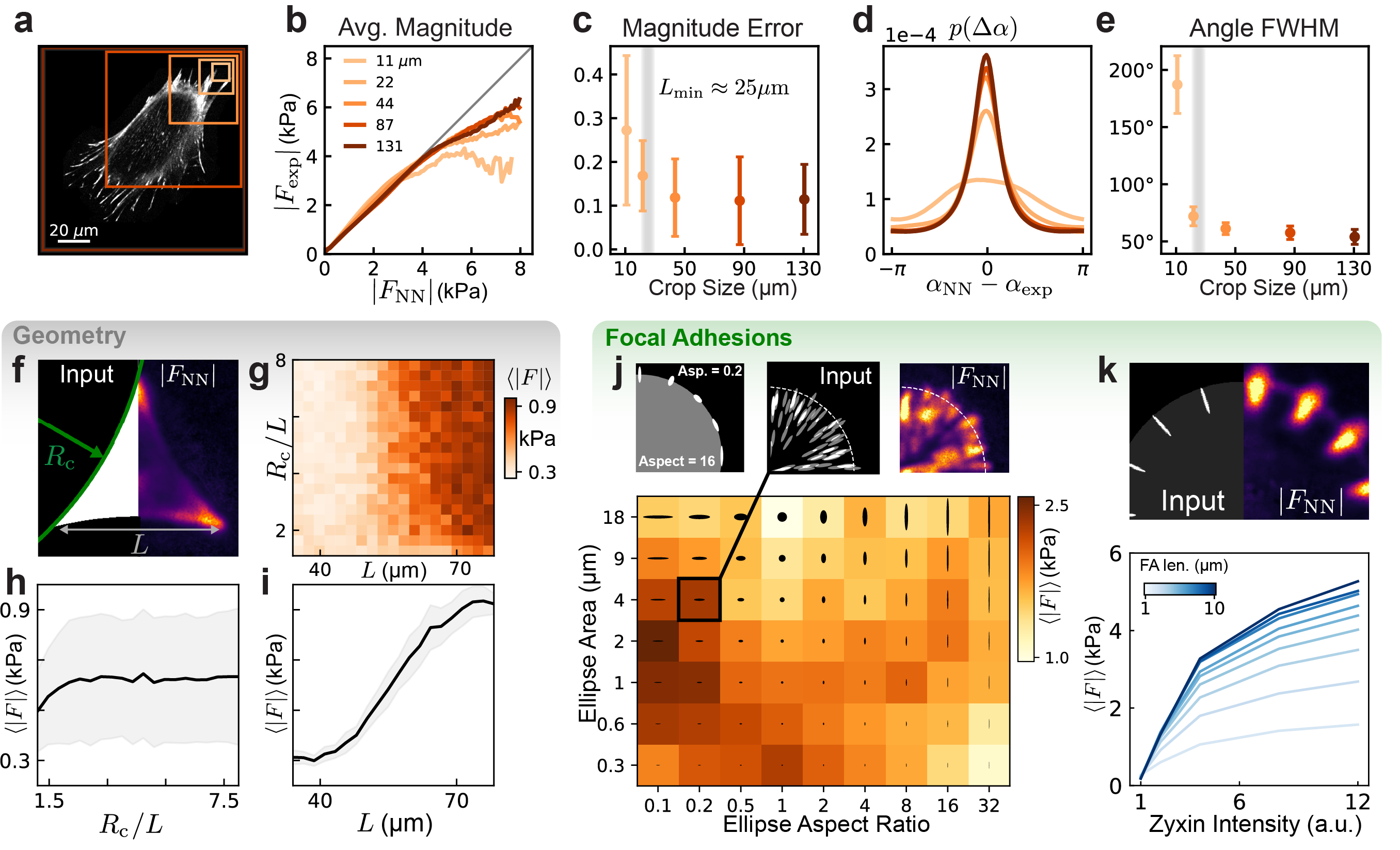}
    \caption{\textbf{Neural networks learn biologically relevant features} 
    \\
    (\textbf{a}) Networks are trained with varying crop sizes, ranging from 64 pixels ($\approx 10 \mu $m) to 768 pixels ($\approx 130 \mu $m).
    (\textbf{b}) Average force magnitude (defined in Fig.~\ref{fig:unet_predictions}) for varying crop sizes.
    (\textbf{c}) Magnitude error as a function of crop size, using the same metric defined in Fig.~\ref{fig:proteins}d.
    (\textbf{d}) Distribution of angular errors $\Delta \alpha = \alpha_{\tnn} - \alpha_{\texp}$ for each crop size. 
    Larger crops cause the distribution to peak sharply about $\Delta \alpha = 0$. 
    (\textbf{e}) Full width half maxima of the distributions in (\textbf{d}) as a function of crop size. FWHM reduces dramatically at a crop size of $\approx\,25\,\mu$m, beyond which it plateaus.
    (\textbf{f}) Synthetic cells of size $L$ consist of three points connected by circular arcs with radius $R_c$. 
    (\textbf{g}) 
    Dependence of average force predicted by mask-trained U-Net on radius of curvature relative to the size of the fake cell, and cell size.
    (\textbf{h}) 
    Averaging along the $x$-axis of (\textbf{g}) shows that average predicted force is independent of of relative radius of curvature.
    (\textbf{i}) 
    Averaging along the $y$-axis of (\textbf{g}) shows that average predicted force increases as a function of cell size. 
    Shaded region in both (\textbf{h}) and (\textbf{i}) denotes one standard deviation.
    (\textbf{j}) 
    (Top) We create fake cells composed of ellipses of varying aspect ratio (defined relative to radial direction) and area, which are randomly distributed in a circular boundary.
    A section of one such cell is shown along with the force magnitudes predicted by the zyxin-trained U-Net. 
    (Bottom) Average predicted force magnitudes vary with aspect ratio and area.
    (\textbf{k}) 
    Additional fake cells are generated of evenly spaced, radially oriented focal adhesions with varying length and intensity. 
    (Top) A section of one such cell is shown along with the force magnitudes predicted by the zyxin-trained U-Net.
    (Bottom) Average predicted force varies with zyxin intensity.
    }
    \label{fig:cropsize_fakecells}
\end{figure*}
\subsection*{Neural networks learn biologically relevant features}
After observing its success at predicting traction forces, we next sought to identify which features of the zyxin distributions the U-Net recognized as relevant for making predictions. 
Zyxin encodes both micron-scale structures, such as FAs ($\sim1$~$\mu$m), as well as cell-scale structures like stress fibers $\sim10$-$100$~$\mu$m (Fig.~\ref{fig:unet_predictions}a). 
To probe how the network interprets these features, we trained U-Nets on random image crops of sizes ranging from 10~$\mu$m up to 130~$\mu$m in our input data (Fig.~\ref{fig:cropsize_fakecells}a).
Even when trained on only a small fraction of the cell, these networks learned models which were accurate on average for both force magnitude (Fig.~\ref{fig:cropsize_fakecells}b-c) and direction (Fig.~\ref{fig:cropsize_fakecells}d-e).
In both of these measurements, improvements in the prediction accuracy was negligible as the input size increased beyond $\sim$25~$\mu$m (Fig.~\ref{fig:cropsize_fakecells}c,e).
This indicates that it is not necessary to know the whole-cell geometry and that accurate predictions can be made by considering a smaller neighborhood around any given point.
\par
Previous work has suggested that both cell morphology \cite{thery2006,oakes2014} and FA distribution \cite{Gardel2008,Han2012,thievessen2013,Liu2014} can impact force generation. 
To understand how the U-Nets interpreted these features, we generated synthetic ``cells''  to systematically vary these features and examine the trained models' response~\cite{murdoch2019pnas}.
To probe the role of cell morphology, we evaluated the mask-trained U-Net on cells that were triangular in shape with a width $L$ and whose edges were arcs with radius of curvature $R_c$(Fig.~\ref{fig:cropsize_fakecells}f). 
While the network did not systematically respond to increases in cell edge curvature (Fig.~\ref{fig:cropsize_fakecells}h), we did find that force production increased with total cell size (Fig.~\ref{fig:cropsize_fakecells}i).
This result is consistent with previous work showing that force generation scales with cell area \cite{oakes2014}, and further demonstrates that the network is sensitive to large-scale features of cell geometry.

To probe the role of FA-like features, we created fake cells composed of elliptical ``FAs'' of varying area and aspect ratio which were distributed randomly throughout a circular cell (Fig.~\ref{fig:cropsize_fakecells}j).
The aspect ratio was defined with respect to the radial direction, allowing us to simultaneously probe the response of the network to both orientation and size of the FA-like structures.
We found that the zyxin-trained U-Net predicted the highest forces for ellipses of area $\sim$2 $\mu$m$^2$ and aspect ratio of $\sim0.1$ (i.e. those pointed radially), consistent with experimental descriptions of FAs~\cite{Prager-Khoutorsky2011}. 
We further investigated the role of FA intensity by creating circular cells with uniformly distributed ellipses of fixed intensity and length along the edge (Fig.~\ref{fig:cropsize_fakecells}k).
Upon increasing the intensity of the ellipses, we found a nonlinear response where the magnitude of the predicted traction forces 
rose sharply at first and continued to grow at a slower rate at higher intensities.
Together these data demonstrate how retrospective analysis on a trained network can reveal features that the neural network finds relevant in making predictions. 
Instead of memorizing complex, uninterpretable features of the training data, the U-Net has identified biological features which allow it to accurately generalize predictions of force generation across cell types and biomechanical states.


\subsection*{Learning adhesion enhances an effective elastic model}

While the U-Net learned rules for predicting forces from zyxin which generalize far beyond the domain on which it was trained, we lack understanding of how the network uses features of the input data to make predictions.
In comparison, previous models inspired by classical continuum theory rely on simple hypotheses allowing for maximum interpretability. 
However, they typically lack the ability to make predictions under wide ranges of cell shapes and distributions of localized FAs~\cite{edwards2011prl, mertz2012, oakes2014, Soine2015, Cao2015, Notbohm2016, Oakes2017,  Hanke2018, Vignaud2021, solowiej-wedderburn2022}.
Here, we demonstrate how to incorporate zyxin into continuum mechanical models using neural networks, thereby learning relationships between proteins and physical parameters which extend the generalizability of physical models.

\begin{figure*}
    \centering
    \includegraphics[width=0.9\textwidth]{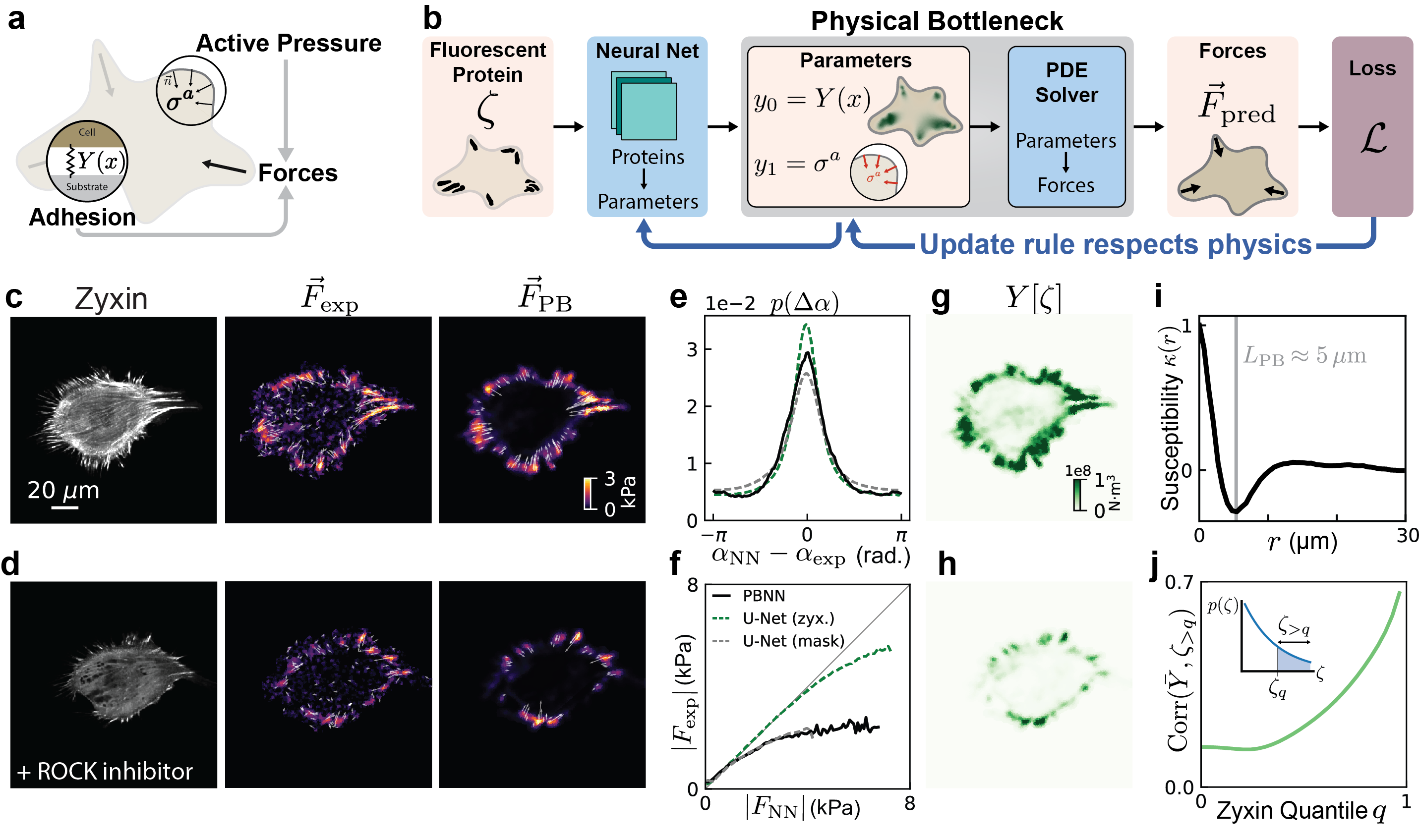}
    \caption{\textbf{Extending physical models using a physical bottleneck}
    \\
    (\textbf{a}) We model the cell as an elastic medium which is subject to a uniform active contractility $\sigma^a$ and is pinned to the substrate with a spatially varying adhesion field $Y(x)$.
    (\textbf{b}) A neural network learns to map zyxin to model parameters, which a PDE solver uses to compute the forces. Network weights are optimized such that physical constraints are always satisfied. We refer to this strong enforcement of the PDE as a ``physical bottleneck'' neural network (PBNN).
    The elastic model with non-uniform adhesion captures forces in cells in both high (\textbf{c}) and low-contractility regimes (\textbf{d}). 
    The PBNN predicts both forces directions (\textbf{e}) and magnitudes (\textbf{f}) more accurately than the full mask-trained U-Net of Fig. \ref{fig:proteins}.
    (\textbf{g-j}) The PBNN reveals how the learned adhesion field depends on zyxin. 
    (\textbf{g}) The learned $Y(x)$ is highly heterogenous and captures the location of FAs.
    (\textbf{h}) The magnitude of the $Y$ field decreases in response to the ROCK inhibition, but remains localized to FAs.
    (\textbf{i}) The ``susceptibility'' of the PBNN, $\kappa(x_i,x_j)=\frac{\partial Y(x_i)}{\partial\zeta (x_j)}$  is sharply peaked within a radius of a few microns.
    (\textbf{j}) We correlate the average adhesion value $\bar Y$ with the amount of zyxin above a threshold $\zeta_q$, where $q$ denotes the $q$th quantile of the zyxin distribution (inset). $\bar{Y}$ correlates strongly only with the highest values of zyxin. 
    }
    \label{fig:physbottleneck}
\end{figure*}

We modeled the cell as an effective two-dimensional active elastic gel adhered to a substrate~\cite{edwards2011prl, mertz2012, oakes2014}.
This model is parameterized by an adhesion strength $Y$ and a global active contractility $\sigma^a$ (Fig. \ref{fig:physbottleneck}a). 
Here we extend this model by considering a spatially-varying adhesion field $Y(x)$, to account for heterogeneous adhesion distributions in the cell~\cite{solowiej-wedderburn2022}.
The forces in this model are calculated as $\vec{F}(x) = Y(x)\vec{u}(x)$ where $\vec{u}(x)$ is the displacement field found by minimizing the system's free energy (see SI for details).
Inspired by the success of the U-Net and to connect chemical quantities to physical model parameters, we made both parameters zyxin-dependent, $Y[\zeta](x)$ and $\sigma^a[\zeta]$, with $\zeta(x)$ denoting the zyxin distribution, so that forces are now given by 
\begin{align}
    \vec{F}(x) = Y[\zeta](x)\vec{u}(x).
    \label{eq: elastic_forces}
\end{align}

While classical methods exist to estimate model parameters from experimental force data, they do not account for the additional constraint that the parameters are functions of zyxin.
To overcome this, we introduce a ``physical bottleneck'' neural network (PBNN) architecture.
The U-Net of Figs. 1-4 calculates forces by processing hundreds of features calculated in the latent layers of the network. 
In contrast, 
our physical bottleneck computes a small number of features from which forces are calculated in a deterministic and well-understood way.
Concretely, the PBNN calculates $Y[\zeta](x)$ and $\sigma^a[\zeta]$ with a neural network and feeds them as parameters into a PDE solver to calculate traction forces~(Fig.~\ref{fig:physbottleneck}c). 
During training, we seek parameters which minimize the mean squared error between predicted forces and the experimentally-measured forces.
In each iterative training step, the adjoint method \cite{troeltzsch} is used to calculate updates to the physical model parameters, which are then passed to the neural network using backpropagation.
This two-step process ensures that updates to the neural network obey the stringent constraints of the physical model.

The PBNN accurately predicts forces and generalizes to strongly perturbed conditions (Fig.~\ref{fig:physbottleneck}c,d).
The predicted force angles~(Fig.~\ref{fig:physbottleneck}e) and magnitudes~(Fig.~\ref{fig:physbottleneck}f), however, are less accurate than those predicted by the zyxin-trained U-Net of Figs. 1-4.
This behavior is expected due to the additional constraints imposed on the PBNN.
The PBNN nevertheless makes more accurate predictions than the mask-trained U-Net~(Fig.~\ref{fig:physbottleneck}e,f). 
This indicates that the two parameters learned at the physical bottleneck contain more relevant information for force prediction than anything an unconstrained deep U-Net could infer from the cell morphology alone.
Moreover, the U-Net processes its latent features with a nearly arbitrarily complex function, while the PBNN processes the $Y$ field and $\sigma^a$ into forces via a simple differential equation.

The introduction of a zyxin-dependent adhesion field $Y[\zeta](x)$ was sufficient to make the physical model competitive with fully deep U-Nets. 
We found that the learned field is strongly heterogeneous and localizes to FA sites~(Fig.~\ref{fig:physbottleneck}g,h).
Furthermore, the intensity of $Y(x)$ decreases on addition of the ROCK inhibitor Y-27632 and mirrors the reorganization and reduction in number of FAs~(Fig.~\ref{fig:physbottleneck}d,h).
However, it is not immediately clear how the PBNN calculated $Y(x)$ from the spatial distribution of zyxin $\zeta(x)$.
To characterize how the adhesion at a point $\mathbf{x}_i$ depends on zyxin at a point $\mathbf{x}_j$ we defined the susceptibility, or linear response, of the network as $\kappa_{\mathbf{x}_i, \mathbf{x}_j} = \frac{\partial Y(\mathbf{x}_i)}{\partial \zeta(\mathbf{x}_j)}$. 
The susceptibility curve exhibits a rapid decay with a minimum at  $\approx$5$\mu$m (Fig.~\ref{fig:physbottleneck}i).
Its shape resembles a Laplace filter commonly used in peak-finding algorithms, indicating that $Y(x)$ is associated with maxima in the zyxin signal.
We further probed the dependence of $Y$ on zyxin by correlating the average adhesion in each image $\bar{Y}$ with the sum of zyxin values above a given threshold (Fig.~\ref{fig:physbottleneck}h, SI Fig.~6). 
Upon increasing the threshold, $\bar Y$ becomes significantly more correlated with zyxin. 
This suggests that the magnitude of the adhesion field is set primarily by the highest zyxin values.
Together, these results indicate that the adhesion field is encoding high-value peaks of zyxin intensity, which correspond to FAs. 

The parameters learned by the PBNN are subject to the assumptions of the model used to constrain them. 
The elastic model makes predictions about displacements within the cell which are not directly accessible experimentally using TFM, nor is it clear what undeformed reference frame these displacements should be measured from. This is due to the fact that a cell, unlike a passive lattice of masses and springs, continuously undergoes
cytoskeletal remodelling even if no external deformations are applied.
Nevertheless, the PBNN is still a powerful tool to test our hypothesized model and it informs us of the minimal necessary ingredients required to predict traction stresses.
In particular, we showed that that cell shape (encoded as boundary conditions), a global contractile ``set-point'' $\sigma^a$, and a field $Y(x)$ encoding focal adhesions were sufficient. 
Furthermore, we have seen that a linear partial differential equation describing an intermediate displacement field is an adequate mathematical model to describe the observed behavior.
\begin{figure*}
    \centering
    \includegraphics[width=\textwidth]{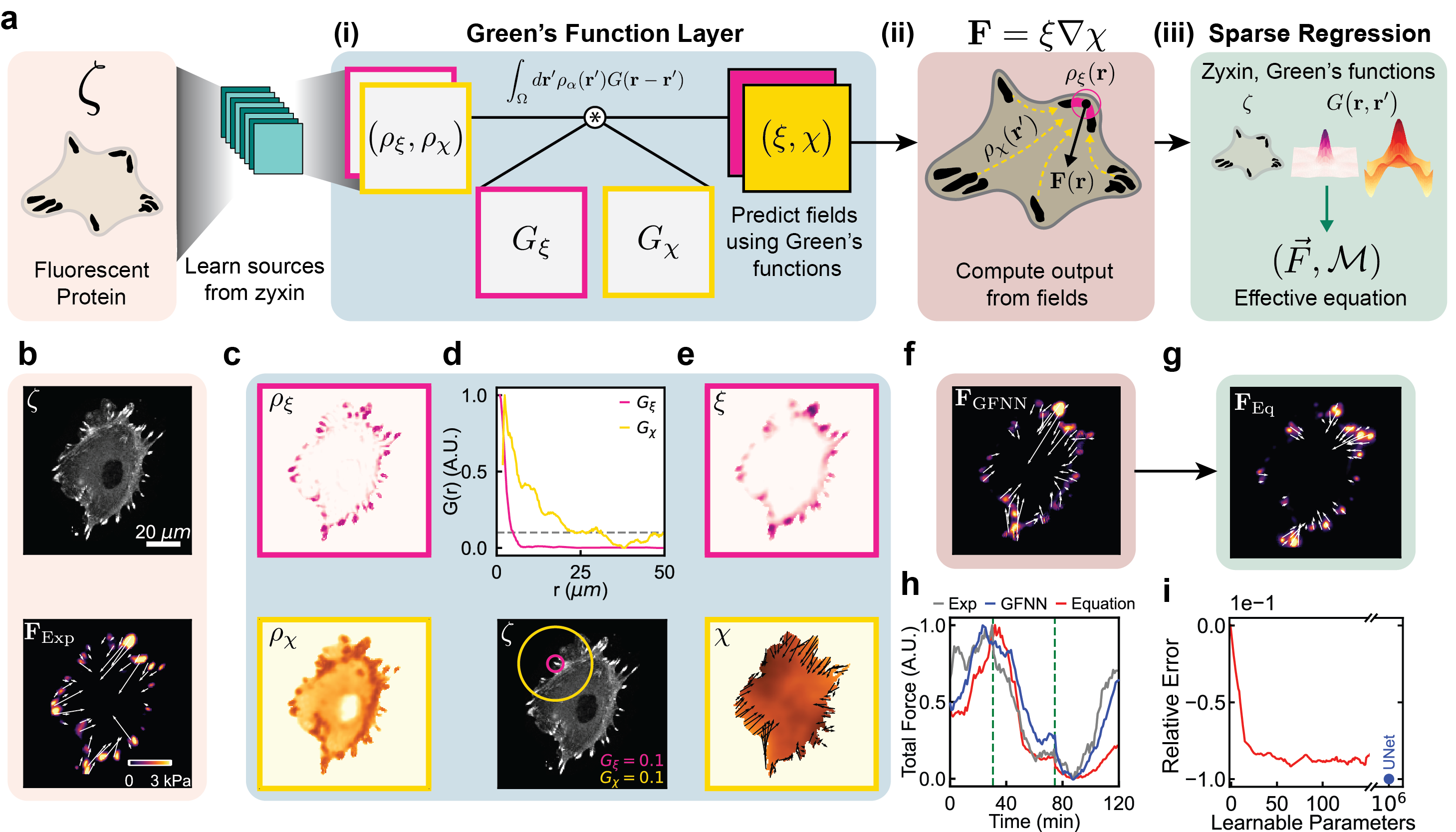}
    \caption{
    \textbf{Model-building pipeline learns physical length scales and effective equations.} \\
    (\textbf{a}) Green's-function neural networks (GFNNs) first extract the fields and long-range interactions needed to predict forces. Next, sparse regression builds effective equations fitting the machine-learned model. 
    (\textbf{b}) The GFNN predicts traction forces from the zyxin intensity field.
    (\textbf{c-e}) The GFNN learns sources $\rho_{\xi}, \rho_{\chi}$ (\textbf{c}) from local zyxin information. These sources are integrated with machine-learned Green's functions (\textbf{d}) to produce the fields $\xi, \chi$ (\textbf{e}). The Green's functions $G_{\xi}, G_{\chi}$ decay over different length scales representing regions over which protein information accumulates (\textbf{d} top). $G_{\xi}$ decays over roughly a focal adhesion size, while $G_{\chi}$ decays more slowly across the cell.
    (\textbf{f}) The predicted force field from the GFNN agrees well with the ground truth (\textbf{b} bottom). 
    (\textbf{g}) Using sparse regression, we learn a formula (see SI) based on the GFNN which predict the force field.
    (\textbf{h}) Time course of predicted forces during a ROCK inhibitor experiment. We compare the experimental forces (grey) to those predicted by GFNN (blue) and the effective equation (red). The dashed lines indicate the drug wash-in and wash-out times.
    (\textbf{i}) Sparse regression yields equations of varying complexity. We plot the improvement in mean-squared error of sparse-regressed models as a function of their complexity, compared to a baseline model $\mathbf{F} = 0$ with no learnable parameters. 
    }
    \label{fig:analyticalmodels}
\end{figure*}

\subsection*{Physics-agnostic model-building reveals relevant length scales}

The success of the PBNN relies on generating plausible hypothesized models, and the insights it produces could be biased by the specific model prescribed.
We now investigate whether we can relax these constraints to gain insights even in the absence of mechanical hypotheses. 
To do this, we turn to a physics-inspired approach to identify machine-learned rules that are agnostic to specific underlying physical models.
This method again trades the complexity of our deep U-Net for fewer, more interpretable operations (Fig.~\ref{fig:analyticalmodels}a). 
Specifically, we assume that the force can be written as a function of machine-learned fields derived from zyxin (yellow and pink boxes in Fig.~\ref{fig:analyticalmodels}a(i)). 
While these fields are analogous to the PBNN's displacement $\vec{u}(x)$ and adhesion $Y(x)$ fields, we do not demand that these quantities obey linear elasticity or any other particular continuum theory. 
We only require that their non-local machine-learned relationships with zyxin density are represented by Green's functions.
The Green's function method is a general tool to calculate a system's response to localized perturbations. For example,
the Green's function of classical electrostatics is the 
$1/r$ potential that determines the effect of a distribution of charges located at a distance $r$.  
With the aid of our machine-learned Green's functions, we will similarly seek to determine how the local traction force depends on zyxin density throughout the cell (Fig.~\ref{fig:analyticalmodels}a(ii)). This is a question for which we do not have the luxury of a readily available formula nor the guarantee \textit{a priori} that an answer even exists.  

Using the same input zyxin images (Fig.~\ref{fig:analyticalmodels}b), we train a Green's function neural network (GFNN)~\cite{gfnn2023} to characterize spatial interactions between our input zyxin images and their respective traction maps. 
The GFNN learns a series of sources and fields (drawn in yellow and pink in Fig.~\ref{fig:analyticalmodels}c-e) from the zyxin images which it uses to predict the traction stresses.
While in principle a GFNN can learn any number of fields, we found that a minimally-complex model could achieve accurate predictions using only two (Fig.~\ref{fig:analyticalmodels}f).
Specifically, the GFNN learned two fields $\xi, \chi$ in terms of which predictions of the traction forces can be made
\begin{align}
    \vec{F} = \xi(x) \vec{\nabla} \chi(x).
    \label{eq: gfnn_forces}
\end{align}
Such a representation is reminiscent of Coulomb electrostatics, with $\xi$ and $\chi$ analogous to the charge and electric potential, respectively.
The ``charge'' $\xi$, in this case, identifies local peaks in zyxin intensity that are similar to focal adhesions (Fig.~\ref{fig:analyticalmodels}c,e top).
The Green's function for $\xi$, $G_\xi$, decays over a very short length scale $\sim 5 \mu$m (Fig.~\ref{fig:analyticalmodels}d), suggesting that it is determined by local information at the adhesion site (Fig.~\ref{fig:analyticalmodels}a(ii),d). 
The ``potential'' $\chi$ is less localized and its Green's function, $G_\chi$, accumulates zyxin information from a larger area of the cell (Fig.~\ref{fig:analyticalmodels}a(ii),c-e bottom). 
This longer decay length suggests that the ``potential'' can infer aspects of the cell morphology from the zyxin distribution. 
Thus, our GFNN model predicts traction forces from interactions between a focal adhesion ``charge'' and a cellular ``potential''.

To simplify this model further, we used sparse regression to build effective equations which approximate the traction forces (Fig.~\ref{fig:analyticalmodels}a(iii)).
Qualitatively accurate analytical formulas can be obtained using only a handful of terms inspired by the GFNN (see SI for full equation). 
Such formulas dramatically compress the information stored in a full U-Net, but they are still capable of predicting traction forces (Fig.~\ref{fig:analyticalmodels}g) and also generalizing to the biochemical perturbations induced by our ROCK inhibition experiments (Fig.~\ref{fig:analyticalmodels}h).  
This formula could capture 60\% of the U-Net predictions, despite containing $10^5$ times fewer parameters~(Fig.\ref{fig:analyticalmodels}i). 
After slightly relaxing the sparsity constraint, a formula with only 25 terms is able to capture 84\% of the U-Net predictions.
Remarkably, this illustrates how the U-Net, a complex black-box, can be distilled into a similarly-accurate formula consisting of two non-local interactions and parameterized by only a handful of terms (Fig.~\ref{fig:analyticalmodels}i).
Our proposed pipeline demonstrates how to extract effective equations which map protein distributions to traction forces without knowing the explicit underlying relations. 
Although no physical input was used to derive them, the structure of Eq. \ref{eq: gfnn_forces} and the learned equation (see SI) are strikingly similar to Eq. \ref{eq: elastic_forces}.
In particular, $\xi$ and $Y$ are both fields which accumulate zyxin information within focal adhesions, while $\vec{\nabla} \chi$ and $\vec{u}$ are vector fields which propagate information throughout the cell.


\section{Discussion}
\begin{wrapfigure}{r}{0.5\textwidth}
    \centering
    \includegraphics[width=0.5\textwidth]{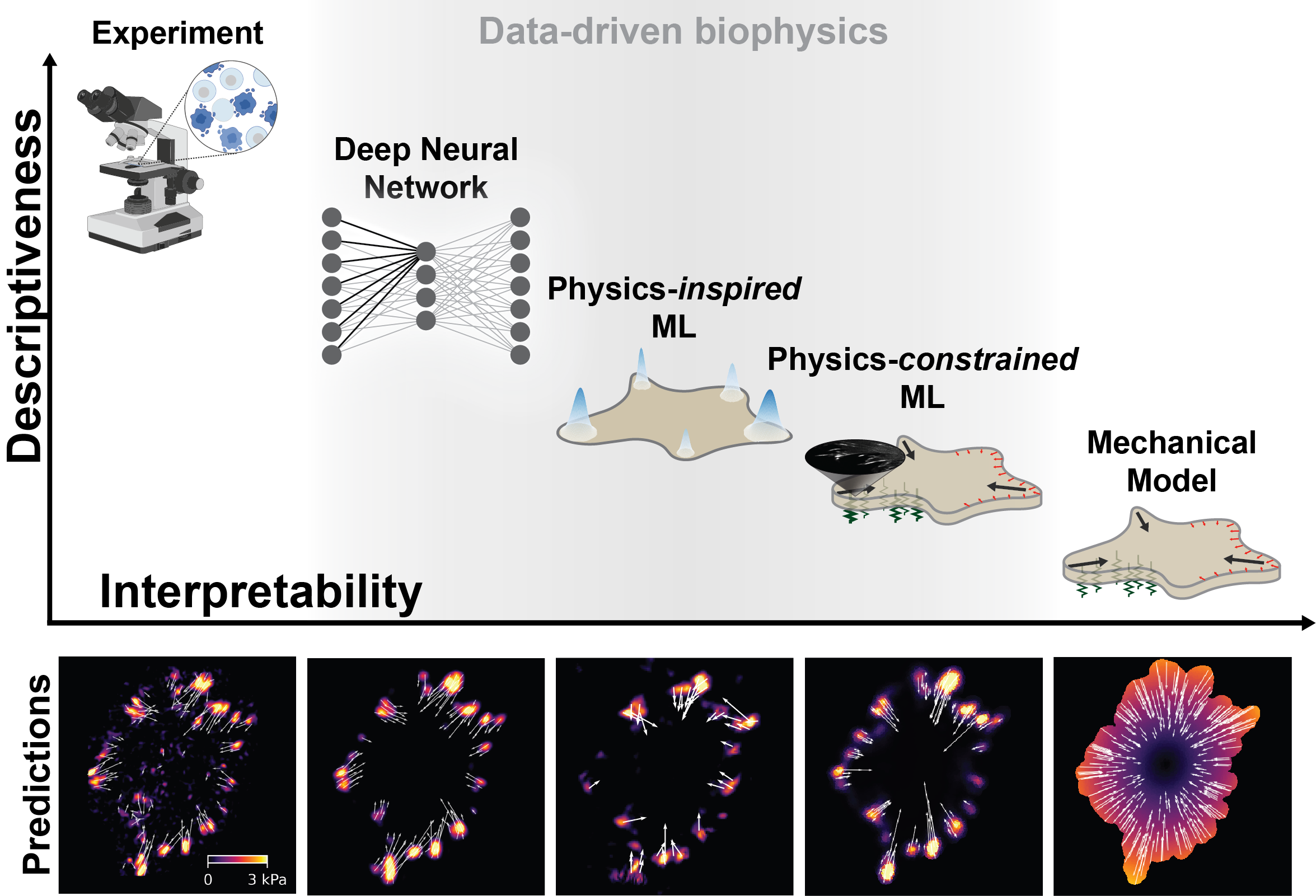}
    \caption{
    \textbf{Machine learning for biophysics}
    \\
    Data-driven biophysical modeling techniques occupy a spectrum of model complexity and interpretability. 
    Deep neural networks are not limited to making black-box predictions, but can serve to test physical hypotheses and build models of various degrees of refinement.
    }
    \label{fig:final_summary}
\end{wrapfigure}
Here, we established that deep neural networks can predict the contractile mechanics of cells directly from images of protein distributions. 
Our results demonstrate that images of a single focal adhesion protein, zyxin, contain sufficient information to accurately predict traction forces. 
We showed that a network trained on images of one cell type collected from one microscope can generalize across a range of cell types, experimental setups, and biomechanical regimes.
These results illustrate the utility of machine learning for extracting robust predictions from heterogeneous, biological data.
Such methods can be implemented with a readily-achievable volume of experimental image data.
This makes them particularly well-suited to predict mechanical behavior
in situations where proteins can be easily imaged but physical measurements are difficult. 


We introduced three data-driven approaches for biophysical modeling which make different trade-offs between descriptiveness and interpretability (Fig.~\ref{fig:final_summary}).
While deep neural networks are not directly interpretable, we demonstrated the utility of synthetic data for identifying relevant components in processes with many interacting proteins.
Our GFNN approach traded some of the complexity of deep U-Nets for interpretable operations, uncovering long-range interactions and even an analytical formula that describes the system behavior.
Finally, we introduced a novel PBNN to test and enhance existing models, which revealed the protein-dependence of effective physical parameters.
These methods represent an alternative approach to hypothesis testing and formulation in the framework of data-driven biophysics (Fig.~\ref{fig:final_summary}).

All three approaches, despite being subject to dramatically different constraints and assumptions, revealed two important length scales.
One length scale of $\sim5\mu$m is consistent with the size of individual focal adhesions and describes the relationship between force magnitude and local zyxin intensity (Fig.~\ref{fig:cropsize_fakecells}j-k; Fig.~\ref{fig:physbottleneck}g-j; Fig.~\ref{fig:analyticalmodels}c-e, top).
Predicting force directions, however, requires information encoded over a larger length scale.
In the GFNN and U-Net, this scale of $\sim 25\mu$m is associated with aspects of cell morphology while in the the PBNN it is accounted for in the PDE's boundary conditions (Fig.~\ref{fig:cropsize_fakecells}c,e; Fig.~\ref{fig:physbottleneck}a-b; Fig.~\ref{fig:analyticalmodels}c-e,bottom). 
Moreover, in the PBNN and GFNN the fields corresponding to long length scales ($\chi,\vec{u}$) and short length-scales ($\xi,Y$) are coupled in a strikingly similar way (Eqs.~\ref{eq: elastic_forces}-\ref{eq: gfnn_forces}).
By using multiple methods which deliver consistent results, we become more confident that the rules derived from our data-driven framework are robust and general. 

The approaches presented here are applicable beyond simple models of cellular contractility. 
Interpretable machine learning methods can lead to an improved understanding of the rules and equations governing spatiotemporal behavior in diverse biological systems.
They may be used to test and extend existing models, as well as learn entirely new ones, in areas where first-principles approaches to biophysics fail.
We only consider prediction of forces from proteins, but an autonomous dynamical model will need to be closed by a relation which predicts how protein distributions evolve in time. 
Our work suggests that it may suffice to consider only the dynamics of an effective adhesion field, rather than accounting for the precise details of cytoskeletal rearrangement.
The methods introduced here could also aid in developing mechano-chemical descriptions of diverse systems such as migrating cells~\cite{Pertz2006migration,ponti2004migration,alert2020migration}, epithelial tissue dynamics~\cite{devany2021PNAS,copenhagen2021topological,saw2017topological} and morphogenesis~\cite{maroudas2021topological,streichan2018eLife}.
These approaches represent a step forward towards harnessing the versatility of machine learning to tackle the complexity of living systems. 

\subsection*{Acknowledgements}
The authors thank M. Fruchart, C. Scheibner, E. Efrati, and M. Han for helpful discussions and suggestions.
M.S.S was supported by the National Science Foundation under Grant No. 2022023.
S.S. was supported by American Heart Association (AHA; Grant number 915248).
M.L.G. and V.V. acknowledge partial support from the UChicago Materials Research Science and Engineering Center (NSF DMR-2011864).  
M.L.G. acknowledges support from the National Institutes of Health (NIH) through awards RO1-GM143792 and RO1-GM104032. 
P.W.O. acknowledges support in part by a National Science Foundation CAREER Award \#2000554 and National Institutes of Health (NIH) National Institute of Allergy and Infectious Disease (NIAID) (Award \# P01-AI02851).
This work was completed in part with resources provided by the University of Chicago’s Research Computing Center.
\end{onehalfspacing}

\newpage

\bibliographystyle{unsrt.bst} 
\textnormal{\bibliography{main.bib}}

\begin{thebibliography}{10}

\bibitem{Pegoraro2017}
Adrian~F Pegoraro, Paul Janmey, and David~A Weitz.
\newblock Mechanical properties of the cytoskeleton and cells.
\newblock {\em Cold Spring Harb. Perspect. Biol.}, 9(11), November 2017.

\bibitem{Blanchoin2014}
Laurent Blanchoin, Rajaa Boujemaa-Paterski, Cecile Sykes, and Julie Plastino.
\newblock Actin dynamics, architecture, and mechanics in cell motility.
\newblock {\em Physiol. Rev.}, 94(1):235--263, January 2014.

\bibitem{Fletcher2010}
Daniel~A Fletcher and R~Dyche Mullins.
\newblock Cell mechanics and the cytoskeleton.
\newblock {\em Nature}, 463(7280):485--492, January 2010.

\bibitem{Svitkina2018}
Tatyana Svitkina.
\newblock The actin cytoskeleton and {Actin-Based} motility.
\newblock {\em Cold Spring Harb. Perspect. Biol.}, 10(1):a018267, January 2018.

\bibitem{Phillips2009}
Robert~Brooks Phillips, Jane Kondev, and Julie Theriot.
\newblock {\em Physical Biology of the Cell}.
\newblock Garland Science, 2009.

\bibitem{MacKintosh2010}
F~MacKintosh and C~Schmidt.
\newblock Active cellular materials.
\newblock {\em Curr. Opin. Cell Biol.}, 22:29--35, August 2010.

\bibitem{Battle2016}
Christopher Battle, Chase~P Broedersz, Nikta Fakhri, Veikko~F Geyer, Jonathon
  Howard, Christoph~F Schmidt, and Fred~C MacKintosh.
\newblock Broken detailed balance at mesoscopic scales in active biological
  systems.
\newblock {\em Science}, 352(6285):604--607, April 2016.

\bibitem{Prost2015}
J~Prost, F~Julicher, and J-F Joanny.
\newblock Active gel physics.
\newblock {\em Nat. Phys.}, 11(2):111--117, February 2015.

\bibitem{Romani2021}
Patrizia Romani, Lorea Valcarcel-Jimenez, Christian Frezza, and Sirio Dupont.
\newblock Crosstalk between mechanotransduction and metabolism.
\newblock {\em Nat. Rev. Mol. Cell Biol.}, 22(1):22--38, January 2021.

\bibitem{Carleo2019}
Giuseppe Carleo, Ignacio Cirac, Kyle Cranmer, Laurent Daudet, Maria Schuld,
  Naftali Tishby, Leslie Vogt-Maranto, and Lenka Zdeborov{\'{a}}.
\newblock {Machine learning and the physical sciences}.
\newblock {\em Reviews of Modern Physics}, 91(4):45002, 2019.

\bibitem{Cichos2020}
Frank Cichos, Kristian Gustavsson, Bernhard Mehlig, and Giovanni Volpe.
\newblock Machine learning for active matter.
\newblock {\em Nature Machine Intelligence}, 2(2):94--103, 2020.

\bibitem{Zaritsky2021}
Assaf Zaritsky, Andrew~R Jamieson, Erik~S Welf, Andres Nevarez, Justin Cillay,
  Ugur Eskiocak, Brandi~L Cantarel, and Gaudenz Danuser.
\newblock Interpretable deep learning uncovers cellular properties in
  label-free live cell images that are predictive of highly metastatic
  melanoma.
\newblock {\em Cell Syst}, 12(7):733--747.e6, July 2021.

\bibitem{Soelistyo2022}
Christopher~J Soelistyo, Giulia Vallardi, Guillaume Charras, and Alan~R Lowe.
\newblock Learning biophysical determinants of cell fate with deep neural
  networks.
\newblock {\em Nature Machine Intelligence}, 4(7):636--644, June 2022.

\bibitem{Jumper2021}
John Jumper, Richard Evans, Alexander Pritzel, Tim Green, Michael Figurnov,
  Olaf Ronneberger, Kathryn Tunyasuvunakool, Russ Bates, Augustin {\v
  Z}{\'\i}dek, Anna Potapenko, Alex Bridgland, Clemens Meyer, Simon A~A Kohl,
  Andrew~J Ballard, Andrew Cowie, Bernardino Romera-Paredes, Stanislav Nikolov,
  Rishub Jain, Jonas Adler, Trevor Back, Stig Petersen, David Reiman, Ellen
  Clancy, Michal Zielinski, Martin Steinegger, Michalina Pacholska, Tamas
  Berghammer, Sebastian Bodenstein, David Silver, Oriol Vinyals, Andrew~W
  Senior, Koray Kavukcuoglu, Pushmeet Kohli, and Demis Hassabis.
\newblock Highly accurate protein structure prediction with {AlphaFold}.
\newblock {\em Nature}, July 2021.

\bibitem{Lin2022biorxiv}
Zeming Lin, Halil Akin, Roshan Rao, Brian Hie, Zhongkai Zhu, Wenting Lu, Nikita
  Smetanin, Robert Verkuil, Ori Kabeli, Yaniv Shmueli, Allan dos Santos~Costa,
  Maryam Fazel-Zarandi, Tom Sercu, Salvatore Candido, and Alexander Rives.
\newblock Evolutionary-scale prediction of atomic level protein structure with
  a language model.
\newblock {\em bioRxiv}, 2022.

\bibitem{Iskratsch2014}
Thomas Iskratsch, Haguy Wolfenson, and Michael~P Sheetz.
\newblock Appreciating force and shape---the rise of mechanotransduction in
  cell biology.
\newblock {\em Nat. Rev. Mol. Cell Biol.}, 15(12):825--833, December 2014.

\bibitem{Murrell2015}
Michael Murrell, Patrick~W Oakes, Martin Lenz, and Margaret~L Gardel.
\newblock Forcing cells into shape: the mechanics of actomyosin contractility.
\newblock {\em Nat. Rev. Mol. Cell Biol.}, 16(8):486--498, August 2015.

\bibitem{Schwarz2012}
Ulrich~S Schwarz and Margaret~L Gardel.
\newblock United we stand: integrating the actin cytoskeleton and cell-matrix
  adhesions in cellular mechanotransduction.
\newblock {\em J. Cell Sci.}, 125(Pt 13):3051--3060, July 2012.

\bibitem{Burridge2015}
Keith Burridge and Christophe Guilluy.
\newblock Focal adhesions, stress fibers and mechanical tension.
\newblock {\em Exp. Cell Res.}, October 2015.

\bibitem{Sabass2008}
Benedikt Sabass, Margaret~L Gardel, Clare~M Waterman, and Ulrich~S Schwarz.
\newblock High resolution traction force microscopy based on experimental and
  computational advances.
\newblock {\em Biophys. J.}, 94(1):207--220, December 2008.

\bibitem{Huang2019}
Yunfei Huang, Christoph Schell, Tobias~B Huber, Ahmet Nihat Amp X0015e
  Imamp~X0015f Ek, Nils Hersch, Rudolf Merkel, Gerhard Gompper, and Benedikt
  Sabass.
\newblock Traction force microscopy with optimized regularization and automated
  bayesian parameter selection for comparing cells.
\newblock {\em Sci. Rep.}, 9(1):1--16, January 2019.

\bibitem{mertz2012}
Aaron~F Mertz, Shiladitya Banerjee, Yonglu Che, Guy~K German, Ye~Xu, Callen
  Hyland, M~Cristina Marchetti, Valerie Horsley, and Eric~R Dufresne.
\newblock Scaling of traction forces with the size of cohesive cell colonies.
\newblock {\em Phys. Rev. Lett.}, 108(19):198101, May 2012.

\bibitem{oakes2014}
Patrick~W. Oakes, Shiladitya Banerjee, M.~Christina Marchetti, and Margaret~L.
  Gardel.
\newblock Geometry regulates traction stresses in adherent cells.
\newblock {\em Biophysical Journal}, 107:825--833, 2014.

\bibitem{Soine2015}
J{\'e}r{\^o}me R~D Soin{\'e}, Christoph~A Brand, Jonathan Stricker, Patrick~W
  Oakes, Margaret~L Gardel, and Ulrich~S Schwarz.
\newblock Model-based traction force microscopy reveals differential tension in
  cellular actin bundles.
\newblock {\em PLoS Comput. Biol.}, 11(3):e1004076, March 2015.

\bibitem{Cao2015}
Xuan Cao, Yuan Lin, Tristian~P Driscoll, Janusz Franco-Barraza, Edna Cukierman,
  Robert~L Mauck, and Vivek~B Shenoy.
\newblock A chemomechanical model of matrix and nuclear rigidity regulation of
  focal adhesion size.
\newblock {\em Biophys. J.}, 109(9):1807--1817, November 2015.

\bibitem{Notbohm2016}
Jacob Notbohm, Shiladitya Banerjee, Kazage J~C Utuje, Bomi Gweon, Hwanseok
  Jang, Yongdoo Park, Jennifer Shin, James~P Butler, Jeffrey~J Fredberg, and
  M~Cristina Marchetti.
\newblock Cellular contraction and polarization drive collective cellular
  motion.
\newblock {\em Biophys. J.}, 110(12):2729--2738, June 2016.

\bibitem{Oakes2017}
Patrick~W Oakes, Elizabeth Wagner, Christoph~A Brand, Dimitri Probst, Marco
  Linke, Ulrich~S Schwarz, Michael Glotzer, and Margaret~L Gardel.
\newblock Optogenetic control of {RhoA} reveals zyxin-mediated elasticity of
  stress fibres.
\newblock {\em Nat. Commun.}, 8:15817, June 2017.

\bibitem{Hanke2018}
Jana Hanke, Dimitri Probst, Assaf Zemel, Ulrich~S Schwarz, and Sarah
  K{\"o}ster.
\newblock Dynamics of force generation by spreading platelets.
\newblock {\em Soft Matter}, 14(31):6571--6581, August 2018.

\bibitem{Vignaud2021}
Timoth{\'e}e Vignaud, Calina Copos, Christophe Leterrier, Mauricio
  Toro-Nahuelpan, Qingzong Tseng, Julia Mahamid, Laurent Blanchoin, Alex
  Mogilner, Manuel Th{\'e}ry, and Laetitia Kurzawa.
\newblock Stress fibres are embedded in a contractile cortical network.
\newblock {\em Nat. Mater.}, 20(3):410--420, March 2021.

\bibitem{edwards2011prl}
Carina~M. Edwards and Ulrich~S. Schwarz.
\newblock Force localization in contracting cell layers.
\newblock {\em Phys. Rev. Lett.}, 107:128101, Sep 2011.

\bibitem{solowiej-wedderburn2022}
Josephine Solowiej-Wedderburn and Carina~M. Dunlop.
\newblock Sticking around: Cell adhesion patterning for energy minimization and
  substrate mechanosensing.
\newblock {\em Biophysical Journal}, 121(9):1777--1786, May 2022.

\bibitem{Hoffman2006}
Laura~M Hoffman, Christopher~C Jensen, Susanne Kloeker, C-L~Albert Wang,
  Masaaki Yoshigi, and Mary~C Beckerle.
\newblock Genetic ablation of zyxin causes {Mena/VASP} mislocalization,
  increased motility, and deficits in actin remodeling.
\newblock {\em J. Cell Biol.}, 172(5):771--782, February 2006.

\bibitem{ronneberger2015unet}
Olaf Ronneberger, Philipp Fischer, and Thomas Brox.
\newblock U-net: Convolutional networks for biomedical image segmentation.
\newblock In Nassir Navab, Joachim Hornegger, William~M. Wells, and
  Alejandro~F. Frangi, editors, {\em Medical Image Computing and
  Computer-Assisted Intervention -- MICCAI 2015}, pages 234--241, Cham, 2015.
  Springer International Publishing.

\bibitem{Yoshigi2005-ri}
Masaaki Yoshigi, Laura~M Hoffman, Christopher~C Jensen, H~Joseph Yost, and
  Mary~C Beckerle.
\newblock Mechanical force mobilizes zyxin from focal adhesions to actin
  filaments and regulates cytoskeletal reinforcement.
\newblock {\em J. Cell Biol.}, 171(2):209--215, October 2005.

\bibitem{stricker2013}
Jonathan Stricker, Yvonne Beckham, Michael~W Davidson, and Margaret~L Gardel.
\newblock Myosin {II-Mediated} focal adhesion maturation is tension
  insensitive.
\newblock {\em PLoS One}, 8(7):e70652, 2013.

\bibitem{Oakes2018}
Patrick~W Oakes, Tamara~C Bidone, Yvonne Beckham, Austin~V Skeeters,
  Guillermina R Ramirez-San Juan, Stephen~P Winter, Gregory~A Voth, and
  Margaret~L Gardel.
\newblock Lamellipodium is a myosin-independent mechanosensor.
\newblock {\em Proc. Natl. Acad. Sci. U. S. A.}, 115(11):2646--2651, March
  2018.

\bibitem{thery2006}
Manuel Théry, Anne Pépin, Emilie Dressaire, Yong Chen, and Michel Bornens.
\newblock Cell distribution of stress fibres in response to the geometry of the
  adhesive environment.
\newblock {\em Cell Motility}, 63(6):341--355, 2006.

\bibitem{Gardel2008}
Margaret~L Gardel, Benedikt Sabass, Lin Ji, Gaudenz Danuser, Ulrich~S Schwarz,
  and Clare~M Waterman.
\newblock Traction stress in focal adhesions correlates biphasically with actin
  retrograde flow speed.
\newblock {\em J. Cell Biol.}, 183(6):999--1005, December 2008.

\bibitem{Han2012}
Sangyoon~J Han, Kevin~S Bielawski, Lucas~H Ting, Marita~L Rodriguez, and
  Nathan~J Sniadecki.
\newblock Decoupling substrate stiffness, spread area, and micropost density: a
  close spatial relationship between traction forces and focal adhesions.
\newblock {\em Biophys. J.}, 103(4):640--648, August 2012.

\bibitem{thievessen2013}
Ingo Thievessen, Peter~M Thompson, Sylvain Berlemont, Karen~M Plevock, Sergey~V
  Plotnikov, Alice Zemljic-Harpf, Robert~S Ross, Michael~W Davidson, Gaudenz
  Danuser, Sharon~L Campbell, and Clare~M Waterman.
\newblock Vinculin-actin interaction couples actin retrograde flow to focal
  adhesions, but is dispensable for focal adhesion growth.
\newblock {\em J. Cell Biol.}, 202(1):163--177, July 2013.

\bibitem{Liu2014}
Yang Liu, Rebecca Medda, Zheng Liu, Kornelia Galior, Kevin Yehl, Joachim~P
  Spatz, Elisabetta~Ada Cavalcanti-Adam, and Khalid Salaita.
\newblock Nanoparticle tension probes patterned at the nanoscale: impact of
  integrin clustering on force transmission.
\newblock {\em Nano Lett.}, 14(10):5539--5546, October 2014.

\bibitem{murdoch2019pnas}
W.~James Murdoch, Chandan Singh, Karl Kumbier, Reza Abbasi-Asl, and Bin Yu.
\newblock Definitions, methods, and applications in interpretable machine
  learning.
\newblock {\em Proceedings of the National Academy of Sciences},
  116(44):22071--22080, 2019.

\bibitem{Prager-Khoutorsky2011}
Masha Prager-Khoutorsky, Alexandra Lichtenstein, Ramaswamy Krishnan, Kavitha
  Rajendran, Avi Mayo, Zvi Kam, Benjamin Geiger, and Alexander~D Bershadsky.
\newblock Fibroblast polarization is a matrix-rigidity-dependent process
  controlled by focal adhesion mechanosensing.
\newblock {\em Nature Cell Biology}, 13(12):1457--1465, December 2011.

\bibitem{troeltzsch}
Fredi Tröltzsch.
\newblock {\em Optimal Control of Partial Differential Equations: Theory,
  Methods and Applications}, volume 112 of {\em Graduate Studies in
  Mathematics}.
\newblock American Mathematical Society, 2000.

\bibitem{gfnn2023}
Discovering models from data: Greens-function neural networks.
\newblock {\em Manuscript in preparation}, 2023.

\bibitem{Pertz2006migration}
Olivier Pertz, Louis Hodgson, Richard~L Klemke, and Klaus~M Hahn.
\newblock Spatiotemporal dynamics of rhoa activity in migrating cells.
\newblock {\em Nature}, 440(7087):1069--1072, April 2006.

\bibitem{ponti2004migration}
A.~Ponti, M.~Machacek, S.~L. Gupton, C.~M. Waterman-Storer, and G.~Danuser.
\newblock Two distinct actin networks drive the protrusion of migrating cells.
\newblock {\em Science}, 305(5691):1782--1786, 2004.

\bibitem{alert2020migration}
Ricard Alert and Xavier Trepat.
\newblock Physical models of collective cell migration.
\newblock {\em Annual Review of Condensed Matter Physics}, 11(1):77--101, 2020.

\bibitem{devany2021PNAS}
John Devany, Daniel~M. Sussman, Takaki Yamamoto, M.~Lisa Manning, and
  Margaret~L. Gardel.
\newblock Cell cycle-dependent active stress drives epithelia remodeling.
\newblock {\em Proceedings of the National Academy of Sciences},
  118(10):e1917853118, 2021.

\bibitem{copenhagen2021topological}
Katherine Copenhagen, Ricard Alert, Ned~S. Wingreen, and Joshua~W. Shaevitz.
\newblock Topological defects promote layer formation in myxococcus xanthus
  colonies.
\newblock {\em Nature Physics}, 17:211--215, 2021.

\bibitem{saw2017topological}
Thuan~Beng Saw, Amin Doostmohammadi, Vincent Nier, Leyla Kocgozlu, Sumesh
  Thampi, Yusuke Toyama, Philippe Marcq, Chwee~Teck Lim, Julia~M Yeomans, and
  Benoit Ladoux.
\newblock Topological defects in epithelia govern cell death and extrusion.
\newblock {\em Nature}, 544(7649):212--216, 2017.

\bibitem{maroudas2021topological}
Yonit Maroudas-Sacks, Liora Garion, Lital Shani-Zerbib, Anton Livshits, Erez
  Braun, and Kinneret Keren.
\newblock Topological defects in the nematic order of actin fibers as
  organization centers of hydra morphogenesis.
\newblock {\em Nature Physics}, 17:251--259, 2021.

\bibitem{streichan2018eLife}
Sebastian~J. Streichan, Matthew~F. Lefebvre, Nicholas Noll, Eric~F. Wieschaus,
  and Boris~I. Shraiman.
\newblock Global morphogenetic flow is accurately predicted by the spatial
  distribution of myosin motors.
\newblock {\em eLife}, 2018.

\bibitem{Hoffman2003-be}
Laura~M Hoffman, David~A Nix, Beverly Benson, Ray Boot-Hanford, Erika
  Gustafsson, Colin Jamora, A~Sheila Menzies, Keow~Lin Goh, Christopher~C
  Jensen, Frank~B Gertler, Elaine Fuchs, Reinhard F{\"a}ssler, and Mary~C
  Beckerle.
\newblock Targeted disruption of the murine zyxin gene.
\newblock {\em Mol. Cell. Biol.}, 23(1):70--79, January 2003.

\bibitem{Hoffman2012-zv}
Laura~M Hoffman, Christopher~C Jensen, Aashi Chaturvedi, Masaaki Yoshigi, and
  Mary~C Beckerle.
\newblock Stretch-induced actin remodeling requires targeting of zyxin to
  stress fibers and recruitment of actin regulators.
\newblock {\em Mol. Biol. Cell}, 23(10):1846--1859, May 2012.

\bibitem{borghi2010}
Nicolas Borghi, Molly Lowndes, Venkat Maruthamuthu, Margaret~L. Gardel, and
  W.~James Nelson.
\newblock Regulation of cell motile behavior by crosstalk between cadherin- and
  integrin-mediated adhesions.
\newblock {\em Proceedings of the National Academy of Sciences},
  107(30):13324--13329, 2010.

\bibitem{Sala2021-wa}
Stefano Sala and Patrick~W Oakes.
\newblock Stress fiber strain recognition by the {LIM} protein testin is
  cryptic and mediated by {RhoA}.
\newblock {\em Mol. Biol. Cell}, 32(18):1758--1771, August 2021.

\bibitem{Butler2002}
James~P Butler, Iva~Marija Toli{\'c}-N{\o}rrelykke, Ben Fabry, and Jeffrey~J
  Fredberg.
\newblock Traction fields, moments, and strain energy that cells exert on their
  surroundings.
\newblock {\em Am. J. Physiol. Cell Physiol.}, 282(3):C595--605, February 2002.

\bibitem{liu2022convnext}
Zhuang Liu, Hanzi Mao, Chao-Yuan Wu, Christoph Feichtenhofer, Trevor Darrell,
  and Saining Xie.
\newblock A convnet for the 2020s, 2022.

\bibitem{loshchilov2017adamw}
Ilya Loshchilov and Frank Hutter.
\newblock Decoupled weight decay regularization, 2017.

\bibitem{mitusch2019dolfinadjoint}
Sebastian~K. Mitusch, Simon~W. Funke, and Jørgen~S. Dokken.
\newblock dolfin-adjoint 2018.1: automated adjoints for fenics and firedrake.
\newblock {\em Journal of Open Source Software}, 4(38):1292, 2019.

\bibitem{Kingma2014AdamAM}
Diederik~P. Kingma and Jimmy Ba.
\newblock Adam: A method for stochastic optimization.
\newblock {\em CoRR}, abs/1412.6980, 2014.

\bibitem{brunton2016PNAS}
Steven~L. Brunton, Joshua~L. Proctor, and J.~Nathan Kutz.
\newblock Discovering governing equations from data by sparse identification of
  nonlinear dynamical systems.
\newblock {\em Proceedings of the National Academy of Sciences},
  113(15):3932--3937, 2016.

\bibitem{Kaptanoglu2022}
Alan~A. Kaptanoglu, Brian~M. de~Silva, Urban Fasel, Kadierdan Kaheman, Andy~J.
  Goldschmidt, Jared Callaham, Charles~B. Delahunt, Zachary~G. Nicolaou,
  Kathleen Champion, Jean-Christophe Loiseau, J.~Nathan Kutz, and Steven~L.
  Brunton.
\newblock Pysindy: A comprehensive python package for robust sparse system
  identification.
\newblock {\em Journal of Open Source Software}, 7(69):3994, 2022.

\bibitem{desilva2020}
Brian de~Silva, Kathleen Champion, Markus Quade, Jean-Christophe Loiseau,
  J.~Kutz, and Steven Brunton.
\newblock Pysindy: A python package for the sparse identification of nonlinear
  dynamical systems from data.
\newblock {\em Journal of Open Source Software}, 5(49):2104, 2020.

\end{thebibliography}

\section{Materials and Methods}

\subsection*{Mammalian expression vectors}
mApple-actin (Addgene plasmid \#54862), mApple-paxillin (Addgene plasmid \#54935), mApple-myosin light chain (Addgene plasmid \#54920) and mito-mGarnet (Addgene plasmid \#104309) vectors were a kind gift from Michael Davidson. EGFP-Paxillin (Addgene plasmid \#15233) was a kind gift from Rock Horwitz.

\subsection*{Cell culture and transfection}
Mouse embryonic fibroblasts (MEFs) stably expressing EGFP-zyxin \cite{Hoffman2003-be,Hoffman2012-zv} were a kind gift of Mary Beckerle’s laboratory (University of Utah, Salt Lake City, UT). Human Osteosarcoma (U2OS) cells were purchased from ATCC (Manassas, VA). MEFs and U2OS cells were cultured in DMEM (MT10013CV, Corning) supplemented with 10\% fetal bovine serum (MT35-010-CV, Corning) and 1\% antibiotic–antimycotic solution (MT30004CI, Corning) at 37°C and 5\% CO2. MDCK cells were cultured in DMEM high glucose supplemented with 2mM L-glutamine and 10\% fetal bovine serum, also at 37°C and 5\% CO2. At 24 h before each experiment, cells were transfected with 5 $\mu$g total DNA using a Neon electroporation system (ThermoFisher Scientific) and plated on polyacrylamide gels for traction force microscopy analysis.

\subsection*{Live cell imaging}
MEFs and U2OS were imaged in Leibovitz’s L-15 medium without phenol red (21083-027, Gibco), 10\% fetal bovine serum (MT35-010-CV, Corning), and 1\% antibiotic, antimycotic solution \\ (MT30004CI, Corning) at 37°C on a Marianas Imaging System (Intelligent Imaging Innovations) consisting of an Axio Observer 7 inverted microscope(Zeiss) attached to a W1 Confocal Spinning Disk (Yokogawa) with Mesa field flattening (Intelligent Imaging Innovations), a motorized X,Y stage (ASI), and a Prime 95B sCMOS (Photometrics) camera. Illumination was provided by a TTL triggered multifiber laser launch (Intelligent Imaging Innovations) consisting of 405, 488, 561, and 637 nm lasers, using a 63X, 1.4 NA Plan-Apochromat objective (Zeiss). Temperature and humidity were maintained using a Bold Line full enclosure incubator (Oko Labs). The microscope was controlled using Slidebook 6 Software (Intelligent Imaging Innovations). Cells were imaged for 2 hrs at 1 min intervals, with typically 5-6 cells being imaged per experiment. When used, a 2X concentration of 5 µM of the ROCK inhibitor Y27632 (10005583, Cayman Chemical Company) in imaging media was added after 30 minutes. After another 45 minutes (i.e. 75 min in total), the drug containing media was replaced with fresh imaging media. 

EGFP-Paxillin-expressing MDCK (G Type II cells, \cite{borghi2010}) were imaged using a Nikon Ti-E Spinning Disk Confocal microscope with a 40x, 1.15 NA WI objective. Images were acquired at 5-min intervals for 4 h using 488 and 642 lasers, and standard filter sets (Em 525/50, Em 700/75) (Chroma Technology, Bellows Falls, VT). Samples were mounted on the microscope in a humidified stage top incubator maintained at 37C and 5\% CO2. Images were acquired using the Andor Zyla 4.2 CMOS camera (Andor Technology, Belfast, UK).

\begin{table}[]
    \centering
    \begin{tabular}{c|c|c|c|c}
         Day & Cell Type & Proteins & Number of Cells & Time Duration  \\
         \hline
         \hline
         1 & MEF & Zyxin & 4 & 180 \\ 
         2 & MEF & Zyxin & 4 & 240 \\ 
         3 & MEF & Zyxin, Actin & 4 & 120 \\ 
         4 & MEF & Zyxin, Actin & 5 & 120 \\ 
         5 & MEF & Zyxin, Paxillin & 4 & 120 \\ 
         6 & MEF & Zyxin, Myosin & 10 & 120 \\ 
         7 & MEF & Zyxin, Paxillin & 7 & 120 \\ 
         8 & MEF & Zyxin, Mitoch. & 7 & 120 \\ 
         9 & U2OS & Zyxin & 5 & 120 \\ 
         10 & U2OS & Zyxin & 12 & 120 \\ 
         11 & MDCK & Paxillin & 25 & 42 \\ %
    \end{tabular}
    \caption{Summary of the datasets considered in this work. We label each dataset by the day on which it was taken. In the case of MDCK, ``Number of Cells'' indicates the number of cell clusters.}
    \label{tab:data}
\end{table}

\subsection*{Traction Force Microscopy Experiments}
Traction force microscopy was performed as described previously \cite{Sabass2008, Sala2021-wa}. Coverslips were prepared by incubating with a 2\% solution of (3-aminopropyl)trimethyoxysilane (313255000, Acros Organics) diluted in isopropanol. Coverslips were washed with DI water 5 times for 10 min and cured overnight at 37°C. Coverslips were incubated with 1\% glutaraldehyde (16360, Electron Microscopy Sciences) in ddH20 for 30 min at room temperature and washed 3 times for 10 min in distilled water, air dried and stored at room temperature. Polyacrylamide gels (shear modulus for MEFs and U2OS cells: 16 kPa—final concentrations of 12\% acrylamide (1610140, Bio-Rad) and 0.15\% bis-acrylamide (1610142, Bio-Rad); shear modulus for MDCK cells: 2.8 kPa with final concentrations of 7.5\% acrylamide and 0.1\% bis-acrylamide) were embedded with 0.04-µm fluorescent microspheres (F8789, Invitrogen) and polymerized on activated glass coverslips for 30 min - 1 h at room temperature. After polymerization, gels were rehydrated for 45 min, treated with cross-linker Sulfo-Sanpah (22589, Pierce Scientific) and photoactivated for 5 min. Polyacrylamide gels were then washed 3 times with PBS and coupled to matrix proteins, rat tail collagen I (for MDCK cells, overnight at 4°C; Corning) or human plasma fibronectin (for MEFs and U2OS cells, 1 h at room temperature; FC010, Millipore). Following matrix protein cross-linking, cells were plated on the gels and allowed to adhere overnight. Cells were imaged the following day. Immediately after imaging, cells were removed from the gel using 0.05\% SDS and a reference image of the fluorescent beads in the unstrained gel was taken.

Analysis of traction forces was performed using code written in Python according to previously described approaches \cite{Sabass2008, Hanke2018, Sala2021-wa}. Prior to processing, images were flat-field corrected and the reference bead image was aligned to the bead image with the cell attached. Displacements in the beads were calculated using an optical flow algorithm in OpenCV (Open Source Computer Vision Library, https://github/itseez/opencv) with a window size of 8 pixels. Traction stresses were calculated using the FTTC approach  \cite{Butler2002,Sabass2008, Huang2019} as previously described, with a regularization parameter of 6.1$\times 10^{-4}$.

\subsection*{Data processing}
Fluorescent images are normalized to have similar values across all cells, for all different proteins considered.
For each cell, we calculate the mean value of the fluorescent signal $f$ within the cell mask, $\mu_{\text{in}}^{\text{cell}} = \langle \langle f(x, t) \rangle_{x\in \text{mask}}\rangle_t$, and the average value of the signal outside the mask $\mu_{\text{out}}^{\text{cell}} = \langle \langle f(x, t) \rangle_{x\notin \text{mask}}\rangle_t$.
The signal is then normalized as $f_{\text{norm}}(x,t) = (f(x,t) - \mu_{\text{out}}^{\text{cell}})/(\mu_{\text{in}}^{\text{cell}} - \mu_{\text{out}}^{\text{cell}})$ and any negative values (corresponding to values below the noise value of empty space) are set to 0. This ensures that $f_{\text{norm}}$ has a mean value of approximately 1.
Cell masks are binary and are generated by thresholding the zyxin channel in each image and filling any holes which appear.

Due to variations in substrate preparation, forces measured by cell depend slightly on the experimental round they belonged to. In our case this corresponds to the day on which they were measured (cf. Fig.~\ref{fig:unet_predictions}).
We therefore normalize the forces of each cell by the average within their dataset, $\mu_F^{\text{day}} = \langle \langle |F(x,t)| \rangle_{x,t}\rangle_{\text{cell}\in\text{day}}$, so that $\vec{F}_{\text{norm}}^{\text{cell}}= \vec{F}^{\text{cell}}/\mu_F^{\text{day}}$ for each cell in day.
Normalized fluorescent signals and forces are used everywhere in this work.

\begin{table}[]
    \centering
    \begin{tabular}{c|c|c|c}
         Network & Figures & Trained on & Evaluated on \\
         \hline
         \hline
         U-Net$^1$ & 1, 4  & (16 cells) D1-6, zyxin & D1: \{2\}, D2: \{1, 2\}, D3:\{2,3,5\}   \\ 
         &&& D4: \{4,5\}, D5: \{2,4\}, D6: \{1, 3,4,5,6\} \\
         & 3  & " & U2OS \\ 
         & 3  & " & MDCK \\ 
         U-Net$^2$ & 2  & (8 cells) D3-4, zyxin & $*$ \\ %
          & 2  & (8 cells) D3-4, mask & $*$ \\ %
          & 2  & (8 cells) D3-4, actin & $*$ \\ %
          & 2  & (8 cells) D3-4, actin and zyxin & $*$ \\ %
          & 2  & (10 cells) D5, 7, zyxin & $*$ \\ %
          & 2  & (10 cells) D5, 7, mask & $*$ \\ %
          & 2  & (10 cells) D5, 7, paxillin & $*$ \\ 
          & 2  & (10 cells) D5, 7, paxillin and zyxin & $*$ \\ 
          & 2  & (9 cells) D6, zyxin & $*$ \\ 
          & 2  & (9 cells) D6, mask & $*$ \\ 
          & 2  & (9 cells) D6, myosin & $*$ \\ 
          & 2  & (9 cells) D6, myosin and zyxin & $*$ \\ 
          & 2  & (6 cells) D8, zyxin & $*$ \\ 
          & 2  & (6 cells) D8, mask & $*$ \\ 
          & 2  & (6 cells) D8, mitoch. & $*$ \\ 
          & 2  & (6 cells) D8, mitoch. and zyxin & $*$ \\ 
    PBNN & 5 & D1: \{3,5\}, D2: \{2,4\}, D3: \{1,3\}, 
    & D1: \{2,4\}, D2: \{3,5\}, D3: \{2,5\},
    \\
    && D4: \{1,2,4\}, D6: \{1\}, zyxin   
    &  D4: \{3,5\}, D6: \{2,3,4,5\},
    \\
    GFNN & 6 & D1 \{1, 2\} D2 \{3\} & D3 \{1\}
    \end{tabular}
    \caption{Overview of the training and testing data used in this work. ``D'' stands for ``Day'', corresponding to datasets in Table \ref{tab:data}. For the protein experiments (U-Net$^2$, marked with an asterisk), separate networks were trained on all but one cell, which was reserved for testing. This was repeated so that each cell in the dataset as the test cell one time.}
    \label{tab:test_train_splits}
\end{table}

\subsection*{U-Net}
Neural networks are implemented in Python using the Pytorch library. Code for network implementation, training, and evaluation is available online (Github). 
To learn force distributions from protein fluorescence images, we exploit the spatial structure of the data by using deep convolutional neural networks.
We use a U-Net architecture \cite{ronneberger2015unet} which is able to represent large-scale features thanks to successive strided convolution layers in an ''encoding`` step which quickly yield a large receptive field that grows exponentially with the number of strided layers. 
At the same time, the presence of skip connections propagates local information forward in the network which prevents the loss of local information in the encoder's coarse-graining.

The channel structure of the U-Net is shown in Table. \ref{tab:unet_struct}. Most layers are composed of blocks with a ``ConvNext'' structure \cite{liu2022convnext}. Briefly, they consist of a layer-wise convolution, batch normalization, an inverse-bottleneck depth-wise convolution, activation function, and finally a depth-wise convolution. Our ConvNext blocks have a layer-wise kernel size of 7 and increase channels in the inverse bottleneck by a factor of 4. For all other convolutions, we use a kernel size of 3. 
Dropout is used with a dropout probability of 10\%.
A detailed illustration of the architecture  is shown in SI~Fig.~1.

The U-Net is trained with the Adam optimizer with weight decay (``AdamW'', \cite{loshchilov2017adamw}) with an initial learning rate of 0.001. The learning rate is scheduled to decay exponentially with rate 0.99. We use a batch size of 8.

For the U-Net used in Figs. 1, 3, and 4, training data consists of 1000 randomly sampled frames from time-lapse series of 16 cells (of 31 cells total).
For the U-Nets used in Figs. 2, training data consists of 600 randomly sampled frames with a variable number of cells for training (see Table \ref{tab:test_train_splits}). 
Each data sample contains an input image (either zyxin, another protein, the mask, or a two-channel zyxin + protein image) paired with the corresponding traction force map measured via TFM. Traction force maps have two channels, which we represent as magnitudes and angles rather than $x$ and $y$ components. 
In all cases, the network is trained for 300 epochs (passes through the entire training data set). 
As a loss function, we take the MSE for the magnitude component, and a $2\pi$-periodic MSE for the angles.

\begin{table}[]
    \centering
    \begin{tabular}{c|c|c|c}
          Layer & Size in & Size out & Details \\
         \hline
         \hline
          Prepended block & $1\times L\times L$ & $C\times L\times L$ 
          & Conv2d, $4\times$ConvNext blocks
          \\
          Skip block 0 & 
          $C\times L\times L$ & 
          $C\times L\times L$  &
          $4\times$ConvNext blocks \\
          Encoder block 0 & 
          $C\times L\times L$ &
          $2C\times \frac{L}{4}\times \frac{L}{4}$ 
          &
          $4\times$ConvNext, BN, Strided Conv2d, GELU \\
          Skip block 1 & 
          $2C\times \frac{L}{4}\times \frac{L}{4}$ &
          $2C\times \frac{L}{4}\times \frac{L}{4}$ 
          & (cf. skip 0)\\
          Encoder block 1 & 
          $2C\times \frac{L}{4}\times \frac{L}{4}$ &
          $4C\times \frac{L}{16}\times \frac{L}{16}$ 
          & (cf. encoder 0)
          \\
          Skip block 2 & 
          $4C\times \frac{L}{16}\times \frac{L}{16}$ & 
          $4C\times \frac{L}{16}\times \frac{L}{16}$ 
          & (cf. skip 0)
          \\
          Encoder block 2 & 
          $4C\times \frac{L}{16}\times \frac{L}{16}$ & 
          $8C\times \frac{L}{64}\times \frac{L}{64}$ 
          & (cf. encoder 0)\\
          Skip block 3 & 
          $8C\times \frac{L}{64}\times \frac{L}{64}$ & 
          $8C\times \frac{L}{64}\times \frac{L}{64}$ 
          & (cf. skip 0)
          \\
          Decoder block 2 & 
          $2C\times \frac{L}{64}\times \frac{L}{64}$& 
          $C\times \frac{L}{16}\times \frac{L}{16} $
          & Upsample, Concat., $4\times$ConvNext, Conv2d \\
          Decoder block 1 & 
          $6C\times \frac{L}{16}\times \frac{L}{16}$  & $2C\times \frac{L}{4}\times \frac{L}{4}$
           & (cf. decoder 2)\\
          Decoder block 0 & 
          $3C\times \frac{L}{4}\times \frac{L}{4}$ & $C\times L\times L$
           & (cf. decoder 2) \\
          Appended block & $C\times L\times L$ & $2\times L\times L$
          &
          $4\times$ConvNext blocks, Conv2d\\
    \end{tabular}
    \caption{Channel structure for the U-Net used in Figs. 1, 3, and 4. We set C=4, while $L$ varies depending on input image size.
    Strided convolutions in the encoder layers have a stride of 4.
    The U-Nets in Fig. 2 are the same, but do not have encoder block 2, skip block 3, or decoder block 2. They also have only 3 ConvNext blocks everywhere instead of 4. 
    }
    \label{tab:unet_struct}
\end{table}

\subsection*{Optimal predictors and histogram plots}
In this work, we evaluate predictions by relying on conditional distributions $p(|F_{\text{exp}}|\big||F_{\text{NN}}|)$. (In the following we consider only force magnitudes and write $|F| = F$, and we abbreviate $F_{\text{NN}}\equiv F_N$ and $F_{\text{exp}}\equiv F_E$ for brevity).
This choice is motivated by the fact that, in the presence of noise, the conditional average $C(F_N) = \mathbb{E}_{F_E} [F_E|F_N]$ will satisfy $C(F_N) =F_N$ for a theoretically optimal predictor and will thus lie along the diagonal in the $F_E - F_N$ plane. 
On the other hand, $C(F_E) = \mathbb{E}_{F_N} [F_N|F_E]$ will generally not lie along this diagonal.

To see this, consider our dataset as a set of pairs $\{X^{(i)}, F^{(i)}\}$ indexed by $i$ where $F_i$ is force magnitude at some pixel, and $X_i$ is the distribution of zyxin in a neighborhood of that pixel. 
The neighborhood is set by the receptive field of the neural network.
Due to either biological or experimental noise, there is a joint (non-deterministic) distribution $p(X, F_E)$ from which our data is drawn. 
The loss function for the force predictions $F_N(X)$ can be written 
\begin{align*}
L(F_N) &= \mathbb{E}_{X}\mathbb{E}_{F_E}\left[(F_N - F_E)^2|X\right] \equiv
\mathbb{E}_{X} L_X(F_N).
\end{align*}
The (Bayes) optimal predictor is one which optimizes, \textit{for every }$X$,
\begin{align*}
    F^{*}_N(X) &= \arg \underset{F_N}{\min}\,\, L_X(F_N).
\end{align*}
In can be shown that $L_X$ is minimized by $F^{*}_N(X)=\mathbb{E}_{F_E} [F_E|X] $. Note that for this to be valid for all $X$, our network must be sufficiently expressive, else our model would be constrained and we would not (necessarily) be able to satisfy this condition for all $X$ independently. If we do indeed have sufficient (infinite) expressivity, this is the (Bayes) optimal predictor.

In the SI, we show that with an optimal predictor, the conditional averages satisfy \\ $C(F_N) = \mathbb{E}_{F_E} [F_E|F_N] = F_N$ and  $C(F_E) = \mathbb{E}_{F_N} [F_N|F_E] = \mathbb{E}_{p(F_E'|F_E)} [F'_E]$. Here $p(F_E'|F_E)$ denotes the posterior predictive distribution $p(F'_E|F_E) = \int dX p(F'_E|X)p(X|F_E)$.
We additionally show that even in the case of Gaussian random variables, the mean of the posterior predictive distribution $p(F'_E|F_E)$ is, for a given $F_E$, \textit{smaller} than $F_E$. Thus even for an optimal predictor, the line defined by $C(F_E)$ lies below the diagonal.
For this reason, we evaluate our predictions by considering $C(F_N)$ and its distance from the diagonal, which is 0 for an optimal predictor.

Probability distributions shown in angle and magnitude plots (for example, Fig. 1e,f) are calculated by binning all pixels of all frames in the test set to calculate the number of joint occurrences of $|\vec{F}_{\text{exp}}|$ and $|\vec{F}_{\text{NN}}|$.
The histogram is normalized to yield a probability and divided by the marginal distribution to calculate conditional probabilities. The ``average'' curves in Figs. 1e, 1f, 2b, 2d, 3b, 3c, 3e, 3f, 4b, 4d, 5e, 5f are given by $C(F_N) = \mathbb{E}_{F_E} [F_E|F_N]$.

\subsection*{Synthetic Cells}
We consider three variants of synthetic cell for the experiments shown in Fig.~4. 
The first variant captures large-scale features of cell geometry. We generate cells of triangles whose sides are given by circular arcs. The cell shape is parameterized by $L$, the distance between the corners of the triangle, and $R_c$, the radius of curvature of the circular arcs.
Forces measured in Fig.~4g,h,i correspond to the average force across the area of the cell.
These synthetic cells were fed as input to a U-Net trained on cell geometry.

The second class of synthetic cells models the distribution of focal adhesion-like objects in the cell. 
In each cell, ellipses of a given aspect ratio and area were randomly distributed (uniformly with a density of 60\%) in a circle of fixed radius of 200 pixels ($\approx 34 \mu$m).
Each cell is parameterized by the corresponding area and aspect ratio of the ellipses.
Each ellipse had an intensity of 1, and they were allowed to overlap. Hence, the input image contained a range of (integer) intensities. 
Ellipse aspect ratio was defined relative to the radial direction, so probing aspect ratio in effect probed focal adhesion orientation.
We evaluate the predicted force by calculating the average force on regions where a focal adhesion is present.

Finally, the role of zyxin intensity was probed by creating cells consisting of equidistant elliptical adhesions on a circular cell ``background''. 
These synthetic cells are parameterized by the intensity of the background $B$, the radius of the cell $R$, the angular density of focal adhesions $D$ ($D=1$ corresponds to no angular space between neighboring adhesions), and the length $L$ and intensity $I$ of focal adhesion ellipses. The intensity of the background models zyxin intensity in the cell away from focal adhesions. 
The zyxin intensity at focal adhesions typically has values in the range 4-12 (a.u.), while the background has values in the range 0-1.
In Fig. 4k, we show the change in intensity for $B=0.8$ and $D=0.5$; results do not strongly depend on $B$ and $D$.
To model the intensity profile of FAs seen in experiment, at the edges of the FA ellipses intensity increases linearly over 2 pixels until the specified FA intensity is reached.
We evaluate the predicted force by calculating the average force on regions where a focal adhesion is present.

Note all values of zyxin here and in the rest of the paper are given in units after normalization described in ``Data processing'' above.

\subsection*{Effective Elastic Model}
We consider a model of the cell as an effective two-dimensional linear elastic medium. 
While originally introduced to model cells on micropillar arrays \cite{edwards2011prl}, 
it has been extended to describe cells uniformly adhered to 2D substrates
\cite{oakes2014}. 
The free energy of the cell is
\begin{align}
    U = \frac{h}{2} \int dA \, (\sigma_{ij}^{\text{el}}  + \sigma^{a}\delta_{ij})u_{ij} + \frac{1}{2} \int dA \, Y(x)u_i u_i
    \label{eq:base_model}
\end{align}
where $u_{ij}=\frac{1}{2}(\partial_i u_j + \partial_j u_i)$ and $\sigma^{\text{el}}_{ij}$ is the elastic stress tensor. $h$ is the height of the cell, which is assumed to be small. As described in the main text, $Y(x)$ models a adhesion or pinning force which penalizes deformations, while $\sigma^a$ serves as an active pressure term.
Minimization of the elastic free energy leads to force balance equations for $\vec{u}(x)$:
\begin{align}
    h \partial_j \sigma_{ij}^{\text{el}} &= Y(x)u_i \quad\text{(in bulk)} \label{eq2:forcebalance}
    \\
     \sigma_{ij}^{\text{el}}n_j &= -\sigma^a n_i \quad\text{(on boundary)}.
\end{align}
In addition to these conditions, we also require that $\sigma^{\text{el}}$ and $\vec{u}$ are related via the constitutive relation
\begin{align}
    \sigma_{ij} &= \frac{E}{1+\nu}\left(\frac{\nu}{1-2\nu}\delta_{ij}u_{kk}+u_{ij}\right), \label{eq1:const}
\end{align}
where $E$ and $\nu$ are the effective Young's modulus and Poisson ratio, respectively, of the cell. Combining the force balance equations with the constitutive relation gives a PDE which determines $\vec{u}$.

\subsection*{Physical Bottleneck}
\begin{table}[]
    \centering
    \begin{tabular}{c|c|c|c}
          Layer & Size in & Size out & Details \\
         \hline
         \hline
          Prepended block & 
          $1\times L\times L$ & $C\times L\times L$ 
          & Conv2d
          \\
          Skip block & 
          $C\times L\times L$ & 
          $C\times L\times L$  &
          10 ConvNext blocks \\
          Encoder block & 
          $C\times L\times L$ & 
          $2C\times \frac{L}{4}\times \frac{L}{4}$ 
          &
          BN, Strided Conv2d, ReLU \\
          Skip block 1 & 
          $2C\times \frac{L}{4}\times \frac{L}{4}$ & 
          $2C\times \frac{L}{4}\times \frac{L}{4}$ 
          & 10 ConvNext blocks \\
          Decoder block & 
          $3C\times \frac{L}{4}\times \frac{L}{4}$ & 
          $1\times L\times L $ 
           & Upsample, Concat., Conv2d 
           \\
           \hline
        Strided Conv2d &
          $1\times L\times L$ & 
          $16 \times \frac{L}{16}\times \frac{L}{16}$  &
          Followed by flattening \\ 
          FC & $\cdot L^2/16$ & N & \\
          FC & N & N & Layer repeated 10 times \\ 
          FC & N & 2 &
    \end{tabular}
    \caption{Channel structure for the physical bottleneck neural network. The top section describes the network used to predict the field $Y(x)$. Here, $C=32$, all Conv2d layers have a kernel size of 5, and ConvNext blocks have kernel size of 15 and inverse bottleneck factor of 4. GELU is used as the activation function throughout.
    The bottom section describes the fully-connected network used to predict the constants $\sigma^a$ and $\nu$. We use $N=32$. In this network, every layer is followed by a ReLU activation.
    }
    \label{tab:pbnn_struct}
\end{table}

The physical bottleneck consists of a neural network step joined with a PDE-solver step.
The neural network is implemented in the PyTorch library, and the PDE-solver is implemented with the dolfin-adjoint library \cite{mitusch2019dolfinadjoint}.
At each step during training, we first predict a field $Y(x)$ and scalars $\sigma^a$ and $\nu$ (the Poisson ratio, which we find to be nearly constant $\nu\approx -1$) using a neural network with zyxin as input. 
The convolutional neural network used to calculate $Y$ is a shallow U-Net with structure shown in Table \ref{tab:pbnn_struct}.
The network used to calculate the scalars consists of one convolutional layer which aggressively coarse-grains the image by a factor of 16, followed by fully-connected layers (see Table \ref{tab:pbnn_struct}).

The parameters output by the neural networks are mapped to a mesh (for spatially-varying parameters) after which they are fed as inputs to a PDE solver.
To solve both forward PDE problems and derive adjoints (described in the following), we use the dolfin-adjoint library \cite{mitusch2019dolfinadjoint}.
The PDE solver calculates a displacement field $\vec{u}(x)$ satisfying the PDE imposed by the physical model. Forces are calculated as $Y(x)\vec{u}(x)$ and compared to the experimentally measured values to give the loss $\mathcal{L}$, which is simply the mean-squared error. 
Gradients $\partial\mathcal{L}/\partial Y(x)$ etc. are computed using the adjoint method. 

We briefly introduce the adjoint method \cite{troeltzsch}, a widely-used technique to optimize PDE parameters in control or data-assimilation tasks.
We consider a PDE which acts on a field $u(x)$ and has parameters $p(x)$.
One wants to optimize a function of the PDE's solution $J(u)$. This can be cast as a constrained optimization problem where one wants to minimize the Lagrangian
\begin{align*}
    \mathcal{L}(u,v,p) = J(u) + \langle v, \mathcal{D}u\rangle.
\end{align*}
Here $\mathcal{D}$ denotes the PDE we wish to optimize (which depends on $p(x)$) and $v(x)$, introduced as a Lagrange multiplier to enforce that $u$ satisfies $\mathcal{D}u=0$, is called the adjoint state.
The angled brackets denote an inner product on 
the function space in which $u$ and $v$ live.
Gradients of the Lagrangian $\partial\mathcal{L}/\partial p$ are given in terms of $v$, which is itself found by solving the adjoint PDE $\mathcal{D}^* p = f(u)$. The adjoint PDE is determined from the Euler-Lagrange equation $\partial\mathcal{L}/\partial u = 0$. 

In practice, the adjoint equations are solved using automatic differentiation. We use dolfin-adjoint to calculate $\partial\mathcal{L}/\partial Y(x)$, $\partial\mathcal{L}/\partial \sigma^a$ and $\partial\mathcal{L}/\partial \nu$. 
These gradients are passed directly to PyTorch's autograd library to update the neural networks which predict $Y(x)$, $\sigma^a$ and $\nu$.
\subsection*{Green's Function Neural Networks}

Green's Function Neural Networks (GFNN) were implemented using the Pytorch Library. 
To predict traction forces with a GFNN, we used the Clebsch decomposition $\vec{F}_{\text{NN}} = \nabla \phi + \xi \nabla \chi$, which is possible for any vector field.
We hypothesized that each Clebsch variable was the solution to a linear partial differential equation (PDE) whose source was a function of the local zyxin density.
\begin{align}
    \mathcal{D}_{\phi} \phi = \rho_{\phi}\left[\zeta\right]; \quad \mathcal{D}_{\xi} \xi = \rho_{\xi}\left[\zeta\right]; \quad \mathcal{D}_{\chi} \chi = \rho_{\chi}\left[\zeta\right]
    \label{eq:funchypothesis}
\end{align}
To predict traction forces subject to this hypothesis, we trained a GFNN to compute each Clebsch variable. Under (\ref{eq:funchypothesis}), each term is the integral of a source and a Green's function.
\begin{align}
    \phi(\vec{x}) &= \int d^2 \vec{r}\, G_{\phi}(\vec{x} - \vec{r}) \rho_{\phi}(\vec{r}); \quad
    \xi(\vec{x}) &= \int d^2 \vec{r}\, G_{\xi}(\vec{x} - \vec{r}) \rho_{\xi}(\mathbf{r}); \quad
    \chi(\vec{x}) &= \int d^2 \vec{r}\, G_{\chi}(\vec{x} - \vec{r}) \rho_{\chi}(\mathbf{r})
\end{align} 

For the network presented in Fig.~\ref{fig:analyticalmodels}, we trained on three cells imaged under normal conditions and evaluated on an unseen cell to which the ROCK inhibition had been applied. 
The inputs were the zyxin density and the outputs were traction force predictions from U-Net discussed in Fig.~\ref{fig:unet_predictions}-\ref{fig:cropsize_fakecells}. 
We center-cropped each input-output pair to a box of size 1024x1024 pixels,  downscaled by a factor of 4, and applied a Fourier cutoff with $k_{\text{max}} = 50$. 
The GFNN used the zyxin field as input and learned to predict forces through the Clebsch decomposition and its corresponding Green's functions.
It predicted the sources $\rho_{\alpha}$ using a shallow convolutional neural network and represented the Fourier-transformed Green's function as a three-channel $256\times256$ complex float tensor ($N=256$ matches the downscaled images size in pixels). The complete network structure is shown in Table~\ref{tab:gfnn_arch}. We trained the network for 200 epochs with batch size 8, learning rate $\lambda=10^{-2}$ on the Green's functions and $\lambda=10^{-4}$ on all other parameters. We used the Adam optimizer~\cite{Kingma2014AdamAM} and scheduled the learning rate to decrease by a factor of 10 whenever the loss function failed to improve for 10 epochs. The GFNN learned to minimize the following loss function with $\beta = 0.1$
\begin{align}
    \mathcal{L} = \sum (\vec{F} - \vec{F}_{GFNN})^2 + \beta \sum |G^{\alpha}(\vec{q})|^2
\end{align}

After training, we found that the $\phi$ field contributed minimally to the predictions. In the Supplementary Text, we demonstrate that the $\nabla \phi$ term in the Clebsch representation accounts for $1.1\%$ of the overall traction force field and is not necessary for the GFNN to generalize to experimental perturbations. Because of this, we omitted it from our analysis in Fig.~\ref{fig:analyticalmodels}.

\begin{table}[]
    \centering
    \begin{tabular}{c|c c | c}
         Module & Layer & Channels & Details \\
         \hline
         \hline
         &  Conv2d & $1\rightarrow64$ & k3$\times$3, groups=64 \\
         Block 1 & Conv2d & $64 \rightarrow 256$ & k1$\times$1 \\
         & Sine & & Activation function \\
         & Conv2d & $256 \rightarrow 64$ & k1$\times$1 \\
         \hline
         & Conv2d & $64\rightarrow64$ & k3$\times$3, groups=64 \\
         Block 2 & Conv2d & $64 \rightarrow 256$ & k1$\times$1 \\
         & Sine & & Activation function \\
         & Conv2d & $256 \rightarrow 64$ & k1$\times$1 \\
         \hline 
         Sources & Conv2d & $64 \rightarrow 3$ & k1$\times$1 \\
         \hline
         & \multicolumn{2}{c|}{FFT2} & \\
         Integration & \multicolumn{2}{c|}{Green's functions} & $\{ \phi, \xi, \chi \} (\mathbf{q}) = G_i(\mathbf{q}) \cdot  f_i (\mathbf{q})$\\
         & \multicolumn{2}{c|}{IFFT2} & \\
         \hline
         Output & \multicolumn{2}{c|}{Clebsch} & $\mathbf{F} = \nabla \phi + \xi \nabla \chi$ 
    \end{tabular}
    \caption{GFNN architecture for cell force prediction. The network includes convolutional blocks inspired by the ConvNext architecture. Grouped convolutions accumulate local information within each channel, while $1\times1$ convolutions with the inverse-bottleneck structure enable the network to learn complex local functions at each pixel while maintaining a minimal receptive field.}
    \label{tab:gfnn_arch}
\end{table}

\subsection*{Sparse Regression}

We performed sparse regression~\cite{brunton2016PNAS} in Python using the PySINDy library~\cite{Kaptanoglu2022,desilva2020}. 
Our candidate library was informed by the Green's function neural network results. We assumed that the sources $\rho_{\alpha}$ were expressible as linear combinations of local zyxin gradients and approximated $G_{\alpha}$ using a set of radially-decaying functions. 
We used a set of local scalar derivatives $\rho_i \in \{ \zeta, \nabla^2 \zeta, (\nabla \zeta)^2, \zeta^2, \zeta \nabla^2 \zeta, \zeta (\nabla \zeta)^2 \}$ and chose the following candidate functions for the Green's functions. 
\begin{align}
    G_i(r) \in \left\{ r^{-1}, \log(r), r, e^{-\ell_{\xi}}, e^{-\ell_{\chi}} \right\} 
\end{align}
The last two terms are exponentially decaying functions whose length scales $\ell_{\alpha}$ were fit to the machine-learned $G_{\xi}, G_{\chi}$ in Fig.~\ref{fig:analyticalmodels}d. 
We included the remaining three terms as slowly-decaying functions which might capture the long-range behavior of $G_{\chi}$.
From the set of sources $\rho_i$ and Green's functions $G_i$, we constructed a library such that $\vec{F}$ could be represented as a linear combination of the following terms.
\begin{multline}
    \vec{F}\left( \vec{x} \right) = \underbrace{\sum_{ij} \left[ { w^{\phi}_{ij}} \,\nabla \int d\vec{r}\, G_i\left(|\vec{x} - \vec{r}|\right) \rho_j \left( \vec{r} \right)    \right] }_{\nabla \phi} + \\
    \underbrace{\sum_{ijk\ell} \left[ {w_{ijk\ell}^{\xi \chi} } \left( \int d\vec{r}\, G_i\left(|\vec{x} - \vec{r}|\right) \rho_j \left( \vec{r} \right) \right) \nabla \left( \int d\vec{r}\, G_k\left(|\vec{x} - \vec{r}|\right) \rho_{\ell} \left( \vec{r} \right) \right) \right]}_{\xi \nabla \chi}
    \label{eq:sparseweights}
\end{multline}
The weight grouping for $\xi \nabla \chi$ (\ref{eq:sparseweights}) is necessary as sparse regression is framed as a \textit{linear} optimization problem.
To obtain the weights $\vec{w}$, we used an elastic net objective. 
\begin{align}
    \vec{w} = \text{argmin} \left[ \langle \big( \vec{F} - \vec{F}(\vec{w}) \big)^2\rangle + \alpha \lvert \lvert \vec{w} \rvert \rvert_1 + \frac 12 \alpha \lvert \lvert \vec{w} \rvert \rvert^2 \right] 
    \label{eq:elasticnet}
\end{align}
Here, $\alpha$ is a parameter which sets the level of solution complexity, which we set to $\alpha=0.5$. This yielded an effective equation with 10 terms (see SI). Choosing a different $\alpha$ yields equations of different complexity. For Fig.~\ref{fig:analyticalmodels}i, we fit formulas using eleven values of $\alpha$ in the range $[10^{-5}, 1]$. We performed this procedure for each cell in the dataset and recorded the number of terms in the resulting formula and the mean-squared error (MSE) with experiment. As a baseline, we also recorded the MSE of the U-Net with experiment. To quantify how adding terms to the formula improves predictions, we defined the Relative Error metric in Fig.\ref{fig:analyticalmodels}i as $MSE_{\text{SINDy}}(\alpha) - MSE_0$, where $MSE_0$ is the error of a model with zero learnable parameters $\vec{F} = 0$, representing the $\alpha \rightarrow \infty$ limit. Thus, a model which uses more learnable parameters to achieve higher accuracy will have a negative relative error.

\newpage

\end{document}